\documentclass{article}  
\RequirePackage[T1]{fontenc} 
\RequirePackage{amsthm,amsmath,amsfonts} 
\usepackage{longtable} 
\usepackage{graphicx} 
\usepackage{latexsym} 
\usepackage{multirow} 
\usepackage{setspace} 
\usepackage{adjustbox} 
\usepackage[bf]{caption} 
\usepackage{emptypage} 
\usepackage{subfigure} 
\usepackage{float} 
 
\RequirePackage[numbers]{natbib} 
\RequirePackage[colorlinks,citecolor=blue,urlcolor=blue]{hyperref} 

\allowdisplaybreaks[1]

\usepackage{times} 
\usepackage{keyval} 
\usepackage{calc} 

\usepackage[letterpaper]{geometry} 
\renewcommand{\baselinestretch}{1.1} 
 
\usepackage{fancyhdr} 
\newcommand{\papertitle}{Convolution-FFT for option pricing in the Heston model} %
\newcommand{\iu}{\mathrm{i}\mkern1mu}

\numberwithin{equation}{section} 
\theoremstyle{plain} 
\newtheorem{theorem}{Theorem}[section] 
 
\newtheorem{remark}{Remark}[section] 
 
\newtheorem{corollary}{Corollary}[section] 
\newtheorem{critere}{Criterion}[theorem] 
\newtheorem{defn}{Definition}[section] 
 
\newtheorem{proposition}{Proposition}[section] 
\newtheorem{variational inequality}{Variational inequality}[section] 
 
\newtheorem{note}{Note}[section]

\DeclareSymbolFont{symbols}{OMS}{cmsy}{m}{n} 
\DeclareFontFamily{OT1}{pzc}{} 
\DeclareFontShape{OT1}{pzc}{m}{it}{<-> s * [1.10] pzcmi7t}{} 
\DeclareMathAlphabet{\mathbf}{OT1}{cmr}{bx}{n} 
\DeclareMathAlphabet{\mathsf}{OT1}{cmss}{m}{n} 
\DeclareMathAlphabet{\mathit}{OT1}{cm}{m}{it} 
\DeclareMathAlphabet{\mathpzc}{OT1}{pzc}{m}{it} 
\DeclareMathAlphabet{\pazocal}{OMS}{zplm}{m}{n}

\DeclareMathOperator{\sgn}{sign} 
\DeclareGraphicsExtensions{.eps} 
\allowdisplaybreaks[1] 
\usepackage{fancyhdr} 
\begin{document} 
	\rhead{\textit{Nov. 3, 2025}} 
	\lhead{\textit{Gao \& Hyndman}} 
	\chead{\textit{\papertitle}} 

	\title{\papertitle} 
	 
	\author{ 
		Xiang Gao 
		\footnote{  
			Department of Mathematics and Statistics,  
			Concordia University,  
			1455 Boulevard de Maisonneuve Ouest, 
			Montr\'eal, Qu\'ebec, 
			Canada H3G 1M8. 
		} 
		\ and  
		Cody Hyndman\footnotemark[1]\ \footnote{Corresponding Author: cody.hyndman@concordia.ca} 
	} 
	 
	\date{November 3, 2025} 
	 
	\maketitle
        
	\abstract{We propose a convolution–FFT method for pricing European options under the Heston model that leverages a continuously differentiable representation of the joint characteristic function. Unlike existing Fourier-based methods that rely on branch-cut adjustments or empirically tuned damping parameters, our approach yields a stable integrand even under large frequency oscillations. Crucially, we derive fully analytical error bounds that quantify both truncation error and discretization error in terms of model parameters and grid settings. To the best of our knowledge, this is the first work to provide such explicit, closed-form error estimates for an FFT-based convolution method specialized to the Heston model. Numerical experiments confirm the theoretical rates and illustrate robust, high-accuracy option pricing at modest computational cost.} 

	\vspace{5mm}

	\vspace{5mm}

	\renewcommand{\baselinestretch}{1.5}

	\section{Introduction\label{sec:Intro}}

        A variety of numerical integration methods are used to efficiently value complex contracts and calibrate financial models. For option pricing models with a closed-form characteristic function, such as the Heston model, it is natural to formulate valuation directly in the Fourier domain. The computational efficiency of the fast Fourier transform (FFT) makes such integration methods particularly attractive for calibration to large sets of plain vanilla options.  A widely used approach is the FFT method of \citet{carr1999option}, which applies a damping factor to the modified payoff in the log-strike domain.

\citet{heston1993closed} proposed a two-factor asset-pricing model with stochastic volatility, derived the characteristic function in closed form, and provided a semi-closed-form solution for pricing vanilla options. In practice, however, the standard representation of the Heston characteristic function is problematic: discontinuities arising from complex logarithms and branch cuts may lead to unreliable numerical integration. Several theoretical and numerical schemes have been introduced to address this discontinuity problem, including the rotation-counting method of \citet{kahl2005not} and the approaches of \citet{lord2010complex} and \citet{levendorskiui2012efficient}. 
 
The seminal work of \citet{carr1999option} introduced an FFT-based framework for option valuation by applying a damping factor to a modified payoff function, enabling efficient numerical integration in the Fourier domain. While widely used, this approach requires careful tuning of the damping exponent and suffers from boundary effects when the transform is applied to non-periodic payoffs. The COS method of \citet{lord2008fast} improves accuracy by expanding the characteristic function in a Fourier-cosine series, but it also relies on finite-interval truncation and requires analytical expressions for characteristic functions. Both methods depend on continuity properties of the Heston characteristic function, which, as noted by \citet{lord2010complex}, can exhibit branch discontinuities and complex-logarithm ambiguities.

In contrast, we derive a differentiable and numerically stable representation of the joint characteristic function for the Heston model that eliminates the discontinuity problem. This representation yields a smoother Fourier integrand and removes the branch-cut ambiguities associated with the standard Heston formulation. This facilitates a convolution-based FFT (CFFT) approach that operates directly in the log-stock domain rather than the log-strike domain. The resulting formulation avoids damping-parameter tuning, reduces boundary artifacts, and provides a natural probabilistic interpretation by convolving the payoff with the transition kernel of the log-price process. It also decouples damping and shifting from the characteristic function itself, which is advantageous for numerical stability.

We note that the term {\it convolution method} has also been used in \citet{lord2008fast}, but in a different sense. Their CONV approach rewrites the discretized Carr and Madan valuation formula as a convolution in the strike variable under a Lévy-process framework. The convolution method developed here arises directly from the conditional density of the Heston log-price process and expresses the option value as the analytical convolution of the payoff with the transition kernel. This density-based formulation leads to a structurally different Fourier integrand and underpins the error analysis that follows.

We derive explicit analytical bounds for both truncation and discretization errors of our convolution--FFT schemes under the Heston model. These bounds specialize and sharpen the general transform-method error analyses of \citet{lee2004option} and \citet{crocce2017error}, and, to the best of our knowledge, have not previously been worked out explicitly in this Heston-specific FFT setting. In this paper we provide an explicit analytical treatment of the truncation and discretization errors that arise in this setting, and develop a convolution--FFT valuation method that incorporates these results in a numerically stable manner.

The outline of the paper is as follows. In Section~\ref{sec1}, we briefly review the Heston model and provide a differentiable expression for the two-dimensional characteristic function. In Section~\ref{sec2}, we introduce the CFFT method for option valuation and provide an error analysis. We also present an efficient modification involving damping and shifting schemes for the option function, which, as our analysis shows, can substantially reduce boundary error. In Section~\ref{sec3}, we present numerical results. To illustrate the advantages of our method, we compare it both to the semi-closed-form solution of \citet{heston1993closed} and to the FFT method of \citet{carr1999option}. Section~\ref{sec4} concludes and an appendix contains proofs.

	\section{Heston model with characteristic function\label{sec1}} 
	The Black-Scholes-Merton model assumes that volatility is constant over time.  The volatility smile refers to the pattern obtained when plotting implied volatility against strike price under the Black–Scholes model.  The volatility smile demonstrates that implied volatility actually varies with strike price.  Restricted by these assumptions, the Black--Scholes model is unrealistic in capturing key features of asset returns, including the volatility smile and skewness in the return distribution. Many empirical studies indicate that volatility is driven by a mean-reverting stochastic process rather than remaining constant; see \citet{fouque2000derivatives}. Therefore, various stochastic volatility models have been proposed to capture such properties. A popular example is the model of  \citet{heston1993closed}, who derived a semi-closed-form solution for pricing vanilla options. Throughout this section we use the notation $(X_t,V_t)$ for the log-price and variance processes. The Heston model parameters are denoted by $\kappa>0$ (mean reversion rate), $\theta>0$ (long-run variance), $\sigma>0$ (volatility of variance), and $\rho\in(-1,1)$ (instantaneous correlation). The term $\eta_t$ will be used for the drift of the log-price process when convenient.

	\subsection{Heston's stochastic volatility model} 
	We assume the stock price $S_t$ obeys a diffusion process on a filtered probability space $(\Omega, \mathcal{F}, \mathbb{P})$. The filtration $\{\mathcal{F}_t\}_{t\geq 0}$ is generated by two independent Wiener processes satisfying the usual conditions of completeness and right continuity. 
Under the Heston model \cite{heston1993closed}, the log-price $X_t$ and variance $V_t$ satisfy 
\begin{align} 
\mathrm{d}X_t &= \left(r - \tfrac12 V_t\right)\mathrm{d}t + \sqrt{V_t}\,\mathrm{d}W^{(1)}_t, 
\label{hest1}\\ 
\mathrm{d}V_t &= \kappa(\theta - V_t)\mathrm{d}t + \sigma\sqrt{V_t}\,\mathrm{d}W^{(2)}_t, 
\label{hest2} 
\end{align} 
with $\mathrm{d}\langle W^{(1)},W^{(2)}\rangle_t = \rho\,\mathrm{d}t$.

\citet{feller1951two} classifies the boundaries for a one-dimensional parabolic diffusion equation and shows that the stochastic volatility process $V_t$ in equation (\ref{hest2}) has the following properties:
	\begin{itemize} 
		\item[(i)] if $2\kappa \theta \geq \sigma^2$, then zero is unattainable and $V_t > 0$, 
		\item[(ii)] if $2\kappa \theta<\sigma^2$, then zero is a regular, attainable and reflecting boundary, which means that $V_t$ can touch 0, but does not spend time there. 
	\end{itemize} 
	 
	We assume the market price of risk scheme $\tilde\Lambda=\left(\Lambda_1,\Lambda_2\right)$ associated with $\left(W_1, W_2\right)$ satisfies the following condition 
	\begin{equation}\label{risk_condition} 
		\frac{\mu - r}{\sqrt{v_t}} = \rho \Lambda_1 + \sqrt{1-\rho^2} \Lambda_2. 
	\end{equation} 
	and we define an equivalent measure $\mathbb{Q}^\Lambda$ on $\mathcal{F}_t$ by 
	\begin{equation*} 
		\frac{d\mathbb{Q}^\Lambda}{d\mathbb{P}} \Bigg|_{\mathcal{F}_t}= \exp\left(-\frac{1}{2}\int_0^t {\left(\Lambda_1^2 + \Lambda_2^2\right)}ds + \int_0^t\Lambda_1 dW_1(s) + \int_0^t{\Lambda_2} dW_2(s)\right). 
	\end{equation*} 
	We have that $\mathbb{Q}^\Lambda$ is equivalent to $\mathbb{P}$ provided that $\mathbb{E}\left[\frac{d\mathbb{Q}^\Lambda}{d\mathbb{P}} |_{\mathcal{F}_t}\right] = 1$ for all $t\in[0,T]$. Though the market price of risk $\tilde\Lambda$ can be chosen arbitrarily, to obtain a complete Heston model we follow Heston's suggestion and let $\Lambda_1(v_t) = \Lambda \sqrt{v_t}$ for some positive constant $\Lambda$ such that $\Lambda_2$ is uniquely determined by equation (\ref{risk_condition}). Further, by Girsanov's theorem, we define two independent Wiener processes under $\mathbb{Q}^\Lambda$ 
	\begin{equation*} 
		\left\{ 
		\begin{aligned} 
			dW^\Lambda_1(t) &= dW_1(t) + \Lambda \sqrt{v_t} dt,\\ 
			dW^\Lambda_2(t) &= dW_2(t) + \frac{\mu - r -\Lambda\rho v_t}{\sqrt{(1-\rho^2)v_t}} dt, 
		\end{aligned} 
		\right. 
	\end{equation*} 
	which gives the risk-neutral dynamics 
	\begin{equation}\label{Heston_neutral} 
		\left\{ 
		\begin{aligned} 
			&dS_t =  r S_t dt + \sqrt{v_t} S_t \left(\rho d{W}^\Lambda_{1t} + \sqrt{1-\rho^2} d{W}^\Lambda_{2t}\right),\\ 
			&dv_t = \bar\kappa\left(\bar\theta - v_t\right)dt + \sigma \sqrt{v_t}d{W}^\Lambda_{1t}, 
		\end{aligned} 
		\right. 
	\end{equation} 
	where $\bar\kappa = \left(\kappa + \sigma\Lambda\right)$, $\bar\theta = \kappa\theta/\bar\kappa$, provided that $\bar\kappa\neq 0$.

	Define the log-stock process, with initial value $x_0 = 0$ 
	\begin{equation*} 
		x_{t} = \log\left(\frac{S_t}{S_0}\right). 
	\end{equation*} 
	Introducing parameter $\tilde\rho = \left(\rho, \sqrt{1-\rho^2} \right)$ and the joint process $dW_t^\Lambda = \left(dW^\Lambda_1(t), dW^\Lambda_2(t)\right)^\top$, we find the dynamics of $x_t$ is given by 
	\begin{equation*} 
		dx_{t} = \left(r - \frac{1}{2}v_t\right) dt + \sqrt{v_t}\tilde\rho dW^\Lambda_t. 
	\end{equation*} 
	The joint process ${X}_t = \left(x_t, v_t\right)^\top$ is given by 
	\begin{equation}\label{joint_process} 
		d{X}_t = \eta(v_t,t) dt + \sqrt{v_t}\xi dW^\Lambda_t, 
	\end{equation} 
	where 
	\begin{align*} 
		\eta(v_t, t) &= \left( \begin{smallmatrix} r - \frac{1}{2}v_t \\ \bar\kappa\left(\bar\theta - v_t\right) \end{smallmatrix} \right), \text{ and }\xi = \left( \begin{smallmatrix} \rho & \sqrt{1-\rho^2} \\ \sigma & 0 \end{smallmatrix} \right). 
	\end{align*} 
 
A central component of Fourier-based pricing methods is the joint characteristic function of $(X_T,V_T)$ conditional on $(X_t,V_t)$. The classical expression derived by Heston \cite{heston1993closed} is widely used in semi-analytical pricing formulas, but direct numerical implementation may suffer from discontinuities due to complex logarithms and branch cuts \cite{kahl2005not,lord2010complex}. For convolution-based methods these discontinuities can lead to instability. In this subsection we present an equivalent formulation of the characteristic function that is continuous in its arguments and differentiable with respect to the model parameters, making it more suitable for numerical integration.

\begin{defn}\label{defn_characteristic}
  The characteristic function of the joint variable $X_t = (x_t, v_t)^{\top}$ under measure $\mathbb{P}$ with initial state $X = (x, v)^{\top}$ and frequency components $U = (p, q)^\top$, is given by
		\begin{equation}\label{charac} 
			\varphi(U, X, t) =  \mathbb{E}^\mathbb{P}\left[e^{\iu U^\top X_T} | X_t=(x, v)^\top\right], 
		\end{equation}
                with terminal condition $\varphi(U, X, T) = e^{\iu U^{\top} X}$.
	\end{defn} 
	Under a different measure, the form of the characteristic function (\ref{charac}) would be different. Similar to the Black-Scholes model, we obtain the following expression of the Heston call with two probability measures 
	\begin{align*} 
		C_t =& e^{-r\tau}\mathbb{E}^\mathbb{Q}\left[(S_T - K)^+\left|S_t,v_t\right.\right]\\ 
		=& e^{-r\tau}\left(\mathbb{E}^\mathbb{Q}\left[S_T\mathbf{1}_{S_T>K}\left|S_t,v_t\right.\right] - K\mathbb{E}^\mathbb{Q}\left[\mathbf{1}_{S_T>K}\left|S_t,v_t\right.\right]\right)\\ 
		=& S_t\mathbb{E}^\mathbb{Q}\left[\frac{S_T}{F(t,T)}\mathbf{1}_{S_T>K}\left|S_t,v_t\right.\right] - Ke^{-r\tau}\mathbb{E}^\mathbb{Q}\left[\mathbf{1}_{S_T>K}\left|S_t,v_t\right.\right]\\ 
		=& S_t\mathbb{E}^{\mathbb{S}}\left[\mathbf{1}_{S_T>K}\left|S_t,v_t\right.\right] - Ke^{-r\tau}\mathbb{E}^\mathbb{Q}\left[\mathbf{1}_{S_T>K}\left|S_t,v_t\right.\right], 
	\end{align*} 
	where $F(t,T) = e^{r\left(T-t\right)}S_t$ is the forward price, as seen from t, and $\tau=T-t$. We define the measure change from the risk neutral measure $\mathbb{Q}$ to the equivalent martingale measure $\mathbb{S}$ which can be seen as an invariant measurement 
	\begin{equation*} 
		\frac{d\mathbb{S}}{d\mathbb{Q}} = \frac{S_T}{F(t,T)}. 
	\end{equation*} 
	For simplicity, we denote $\mathbb{P}_1 = \mathbb{S}$ and $\mathbb{P}_2=\mathbb{Q}$, under which 
	\begin{align*} 
		P_1(S_T , K) =&\mathbb{P}_1(S_T \geq K),\\ 
		P_2(S_T , K) =&\mathbb{P}_2(S_T \geq K), 
	\end{align*} 
	and the pricing formula becomes 
	\begin{equation}\label{Heston_call} 
		C_t = S_tP_1\left(S_T, K\right) - Ke^{-r\tau}P_2\left(S_T, K\right). 
	\end{equation} 
	According to arbitrage pricing theory, the Heston call option $C\left(S,v,t\right)$ satisfies the following PDE (see \citet{heston1993closed} and \citet{black1973pricing}): 
	\begin{align*} 
		\frac{1}{2}vS^2&\frac{\partial^2 C}{\partial S^2} + \rho \sigma v S\frac{\partial^2 C}{\partial S \partial v} + \frac{1}{2}\sigma^2 v \frac{\partial^2 C}{\partial v^2} + r S\frac{\partial C}{\partial S} + \left[\bar\kappa\left(\bar\theta - v\right) - \sigma\Lambda v\right] \frac{\partial C}{\partial v} + \frac{\partial C}{\partial t} - r C = 0. 
	\end{align*} 
	Due to the similar structure to the Black-Scholes model, $P_1$ and $P_2$ must satisfy the following PDE in terms of $x= \ln \frac{S}{K}$ 
	\begin{equation}\label{Hestion_phi} 
		\frac{1}{2}v\frac{\partial^2 P_i}{\partial x^2} + \rho \sigma v \frac{\partial^2 P_i}{\partial x \partial v} + \frac{1}{2}\sigma^2 v \frac{\partial^2 P_i}{\partial v^2} + \left(r + c_i v\right)\frac{\partial P_i}{\partial x} + \left(a - b_i v\right)\frac{\partial P_i}{\partial v} + \frac{\partial P_i}{\partial t} = 0, 
	\end{equation} 
	where 
	$c_1 = \frac{1}{2}$,$c_2 = -\frac{1}{2}$, $a =\bar\kappa\bar\theta$, $b_1= \bar\kappa + \Lambda \sigma- \rho\sigma$, $b_2= \bar\kappa + \Lambda\sigma$ for $i=1,2$. 
	 
	By the Feynman-Kac representation theorem, the characteristic functions $\varphi_i$ defined by (\ref{charac}) under measures $P_i$ satisfying (\ref{Hestion_phi}) are the unique bounded solutions to the PDE 
	\begin{equation}\label{characteristic2_PDE} 
		\frac{\partial \varphi_i}{\partial t} + \left(r + c_i v\right) \frac{\partial \varphi_i}{\partial x} + \left(a - b_i v\right) \frac{\partial \varphi_i}{\partial v} + \frac{1}{2}v \frac{\partial^2 \varphi_i}{\partial x^2} + \frac{\sigma^2}{2} v \frac{\partial^2 \varphi_i}{\partial v^2} + \rho\sigma v \frac{\partial^2 \varphi_i}{\partial x \partial v} = 0, 
	\end{equation} 
	with boundary condition $\varphi = e^{\iu\left(px + qv\right)}$. 
	 
	The discontinuity problem in Heston's characteristic function has been studied and solved by other authors such as \citet{kahl2005not} using phase rotation counting and \citet{cui2017full} splitting the term that causes the phase shift, however, their solutions are not easy to implement in calibration. Therefore, we propose another simple representation of the joint characteristic function.  The next result provides a differentiable representation of the joint characteristic function under the Heston model. This form avoids discontinuities found in the standard expression and will be used in our convolution–FFT method.

	\begin{theorem}\label{thm_characteristic}(Joint characteristic function)  
The characteristic function of the joint process $X_t = (x_t, v_t)^{\top}$ under $P_i$, with initial condition $X = (x, v)^{\top}$ and Fourier variables $U = (p, q)$, is given by
		\begin{equation}\label{my_kernel} 
			\varphi_i\left(p, q\right) = \exp\left(\iu p \left(x + r\tau\right) + \iu q\left(v + a\tau\right) + \frac{ \gamma + \lambda}{\sigma^2}\left(1 - \zeta\right)v - \frac{\gamma - \lambda}{\sigma^2}a\tau +\frac{2 a}{\sigma^2}\ln \zeta\right), 
		\end{equation} 
		where 
		\begin{align} 
			\gamma =& \sqrt{\sigma^2\left(p^2 - 2\iu c_i p\right) + \left(b_i - \iu \sigma\rho p\right)^2},\label{charac_params1}\\ 
			\lambda =& b_i - \iu \sigma\rho p - \iu \sigma^2 q,\label{charac_params2}\\ 
			\zeta =& \frac{2\gamma}{\gamma + \lambda + (\gamma - \lambda)e^{-\gamma\tau}}.\label{charac_params3} 
		\end{align} 
	\end{theorem} 
	\noindent {\it The proof is given in Appendix \ref{Pf_theorem_thm_characteristic}. }

\begin{figure}[H] 
		\centering 
		\caption{Heston's characteristic function} 
		\vspace*{-0.4cm} 
		\includegraphics[width=1\linewidth]{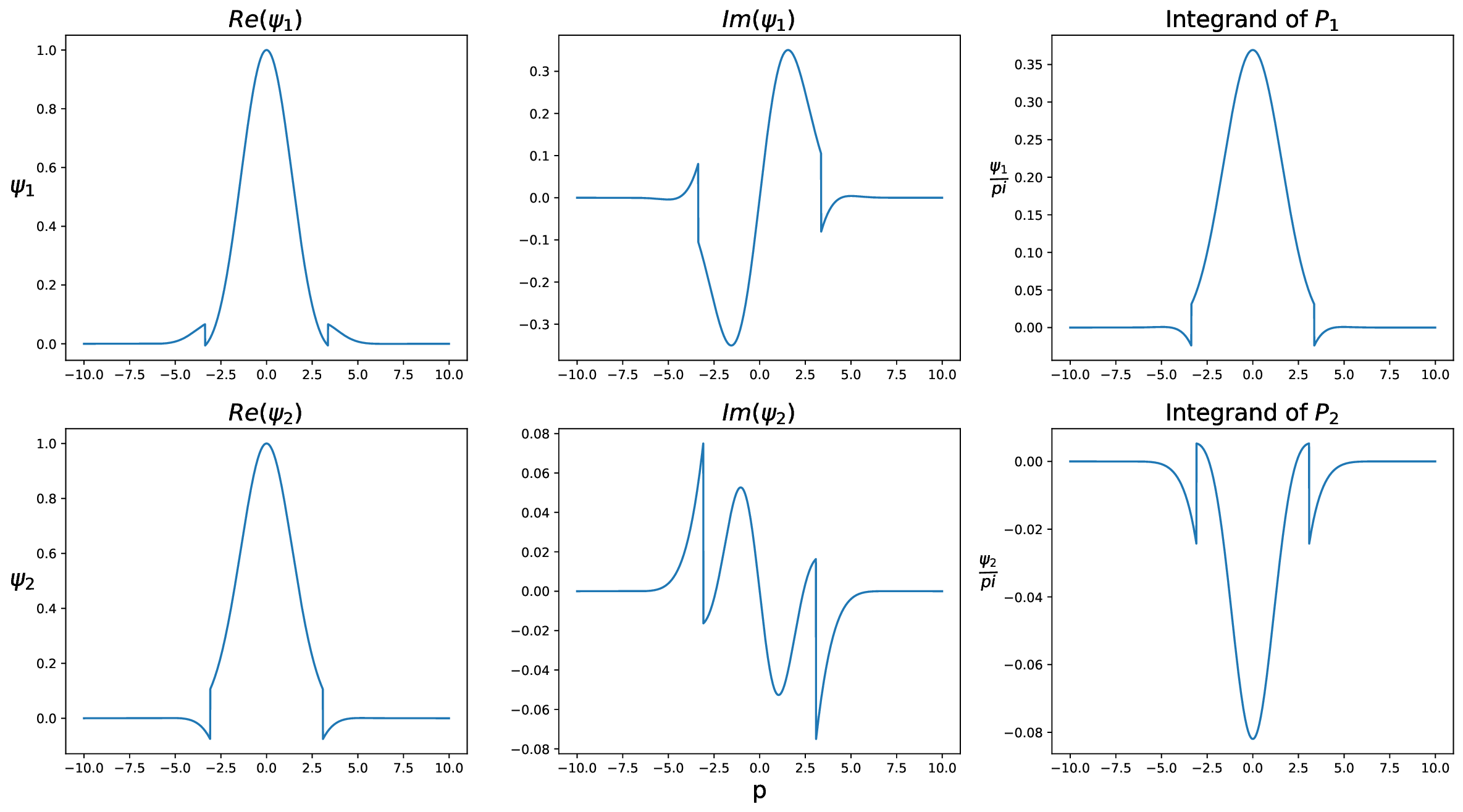} 
		\caption*{$\Lambda = 1$, $r=0.03$, $\rho=-0.8$, $\kappa = 3$, $\theta = 0.1$, $\sigma = 0.25$, $\tau=5$} 
		\label{chap2:pic0_1} %
	\end{figure}

\begin{figure}[H] 
		\centering 
		\caption{Joint characteristic function} 
		\vspace*{-0.4cm} 
		\includegraphics[width=1\linewidth]{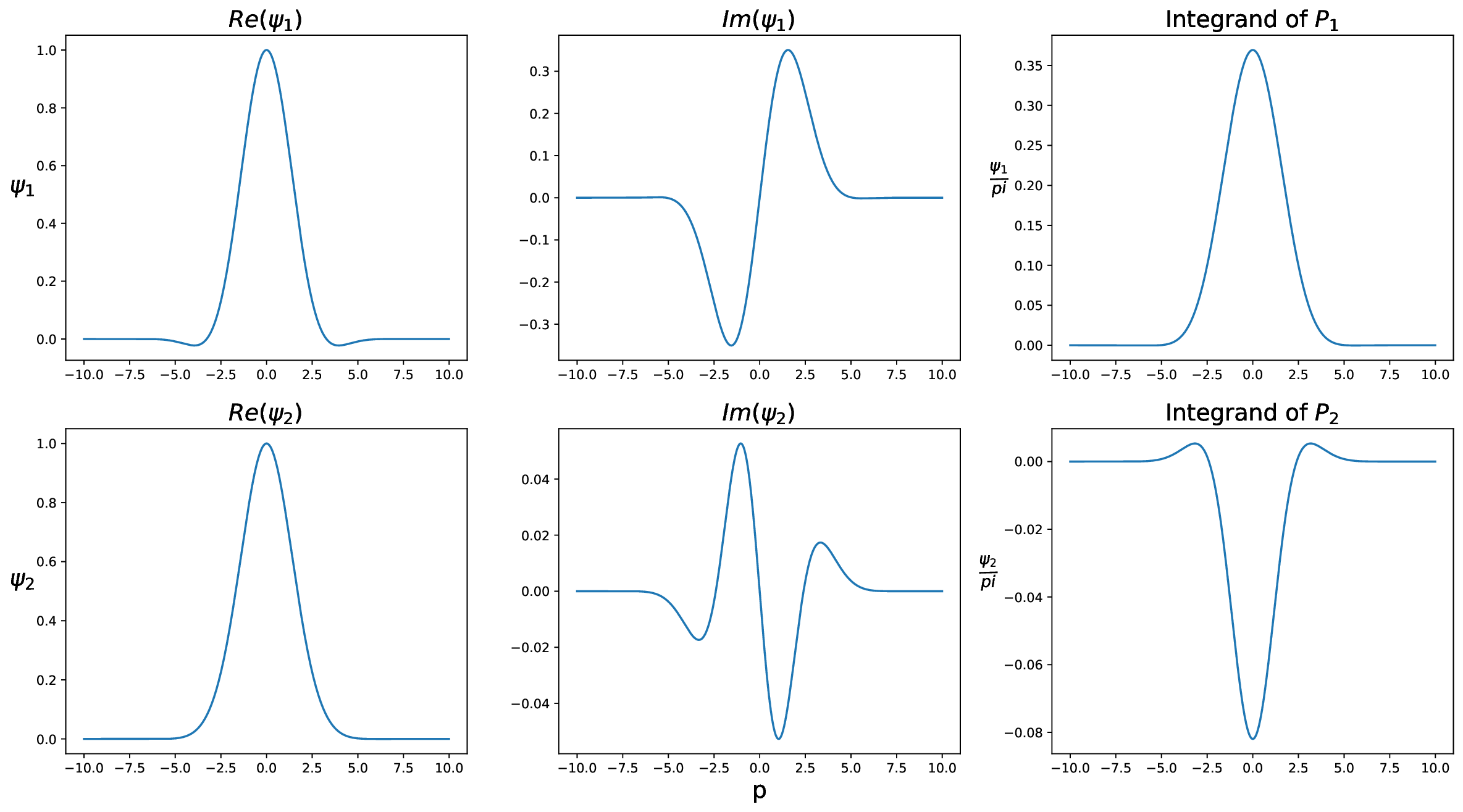} 
		\caption*{$\Lambda = 1$, $r=0.03$, $\rho=-0.8$, $\kappa = 3$, $\theta = 0.1$, $\sigma = 0.25$, $\tau=5$} 
		\label{chap2:pic0_2} %
	\end{figure}

This representation is fully equivalent to the standard form of Heston’s characteristic function, but its continuous differentiability and removal of complex–logarithm ambiguities make it more stable for numerical computation. In particular, this form is advantageous for the convolution–FFT approach developed in Section~\ref{sec2}, where smoothness with respect to the spatial variables plays a key role in controlling truncation and discretization errors. In the estimation, we use the following kernel function obtained from the joint characteristic function of the increment $X_T - X_t$:
	\begin{align}\label{psi} 
		\psi_i(p, q) =& \mathbb{E}\left[ e^{\iu U^\top \left(X_T - X_t\right)}\big|X_t = X\right]
		= e^{-\iu U^\top X}\varphi_i(p, q)\nonumber\\ 
		=& \exp\left(\iu p r \tau + \iu q a\tau + \frac{ \gamma+\lambda}{\sigma^2}\left(1-\zeta\right)v - \frac{\gamma-\lambda}{\sigma^2}a\tau + \frac{2a}{\sigma^2}\ln \zeta\right). 
	\end{align} 
	The characteristic function is used to calculate the values of $P_i$ by letting $q=0$ 
	\begin{align} 
		P_i = \frac{1}{2} + \frac{1}{\pi}\int_0^\infty \text{Re}\left[\frac{\psi_i\left(p\right)}{p\iu}\right]dp.\label{integration} 
	\end{align} 
	The original characteristic function solution given by \citet{heston1993closed} is 
	\begin{equation} 
		\hat\varphi_i(p)=\exp\left\{C\left(T-t, p\right) + D\left(T-t, p\right) + \iu p x\right\}, \label{eq:og_heston_char}
	\end{equation} 
	where 
	\begin{align} 
		C\left(\tau, p\right) &= r p \tau\iu + \frac{a}{\sigma^2}\left\{\left(b_i-\rho\sigma p \iu + \gamma\right)\tau - 2\ln\left[\frac{1-ge^{\gamma r}}{1-g}\right]\right\},\label{logterm}\\ 
		D\left(\tau, p\right) &= \frac{b_i-\rho\sigma p \iu + \gamma}{\sigma^2}\left[\frac{1-e^{\gamma r}}{1-ge^{\gamma r}}\right], \label{logterm2}\\
		g &= \frac{b_i-\rho\sigma p\iu + \gamma}{b_i - \rho\sigma p \iu - \gamma}.\label{logterm3}
	\end{align}
        The term $g$ in equations~\eqref{logterm}-\eqref{logterm3} can encounter a zero denominator, and when this occurs, large values of $p$ cause the argument of the logarithm to rotate rapidly. This effect can be seen from the asymptotic behavior for large $|p|$,
\[
e^{\gamma\tau} \sim e^{\sigma\sqrt{1-\rho^{2}}\,|p|\tau},
\]
which induces rapid phase variation in the logarithm and produces the apparent discontinuity shown in Figure~\ref{chap2:pic0_1}. A detailed discussion of the singularity in the original Heston characteristic function can be found in \citet{MR3375192}. The logarithmic term in the joint characteristic function \eqref{my_kernel} has no singularities. Therefore, the representation in \eqref{my_kernel} does not suffer from the discontinuity problem. Other representations of the characteristic function appear in \citet{kahl2005not}, where an adjustment of the phase rotation is introduced, and in \citet{cui2017full}, where hyperbolic functions are used.   In Figure~\ref{chap2:pic0_1} we present the original characteristic function of \citet{heston1993closed}, and in Figure~\ref{chap2:pic0_2} we present the characteristic function given by \eqref{my_kernel}. The integrands shown in Figures~\ref{chap2:pic0_1} and \ref{chap2:pic0_2} show the values of $\psi_i/(p\iu)$ that appear in the integral \eqref{integration}.

	To obtain a simple expression of the derivative of (\ref{psi}), we introduce the following notation 
	\begin{equation*} 
		\alpha = \frac{\gamma+\lambda}{\sigma^2},~~\beta = \frac{\gamma-\lambda}{\sigma^2}, 
	\end{equation*} 
	and rewrite equation (\ref{psi}) as 
	\begin{equation}\label{simple_varphi} 
		\psi_i(p, q) = \exp\left(\iu p r \tau + \iu q a\tau + \alpha\left(1-\zeta\right) v - \beta a\tau + \frac{2a}{\sigma^2}\ln \zeta\right), \text{ for }~\zeta = \frac{\alpha + \beta}{\alpha + \beta e^{-\gamma\tau}}. 
	\end{equation} 
	We obtain the first-order derivative from (\ref{simple_varphi}) as 
	\begin{align} 
		\frac{\partial \psi_i}{\partial p} =& \varphi\left(\iu r \tau + \left(\alpha_p\left(1-\zeta\right) - \alpha \zeta_p\right)v - \beta_p a\tau + \frac{2a}{\sigma^2} \zeta_1 \right),\label{nabla1}\\ 
		\frac{\partial \psi_i}{\partial q} =& \varphi\left(\left(\alpha_q\left(1-\zeta\right) - \alpha\zeta_q\right)v + \frac{2a}{\sigma^2} \zeta_2\right),\label{nabla2} 
	\end{align} 
	where 
	\begin{align*} 
		\gamma_p &= \frac{\sigma^2(1 - \rho^2)p - \iu(\sigma^2 c_i + \sigma\rho b_i)}{\gamma}, \qquad
		\alpha_p= \frac{\gamma_p - \iu\sigma\rho}{\sigma^2}, \quad\beta_p= \frac{\gamma_p + \iu\sigma\rho}{\sigma^2}, \\
		\zeta_p  &= \frac{\alpha_p + \beta_p}{\alpha+\beta}\zeta - \frac{\alpha_p+\beta_p e^{-\gamma\tau}}{\alpha+\beta}\zeta^2 + \gamma_p\tau \left(1-\frac{\alpha \zeta}{\alpha+\beta}\right)\zeta,\qquad 
		\zeta_q = \frac{1 - e^{-\gamma\tau}}{\alpha+\beta}\zeta^2\iu,\\ 
		\zeta_1 &= \frac{\alpha_p +\beta_p}{\alpha+\beta} - \frac{\alpha_p+\beta_p e^{-\gamma\tau}}{\alpha+\beta}\zeta + \gamma_p\tau \left(1-\frac{\alpha \zeta}{\alpha+\beta}\right), \qquad
		\zeta_2 = \frac{1 - e^{-\gamma\tau}}{\alpha+\beta}\zeta\iu. 
	\end{align*}

        The representation obtained in Theorem~\ref{thm_characteristic} is continuously differentiable in all parameters and avoids the branch–cut discontinuities in the classical Heston characteristic function. This smooth form is particularly advantageous for numerical work, since it ensures stable evaluation of Fourier transforms and facilitates both pricing and calibration. In the next section we incorporate this representation into a convolution–FFT method for option valuation under the Heston model.

	\section{Convolution-FFT method}\label{sec2} 
In this section we apply the differentiable characteristic function obtained in Section~\ref{sec1} to the convolution method for option valuation. The value of a European option can be written as the convolution of the payoff with the transition density of the log-price process. When the convolution is evaluated on a truncated interval using the discrete Fourier transform, the FFT can be used to compute the option value efficiently on a uniform spatial grid. The use of the smooth characteristic function developed above improves the numerical stability of these Fourier-based calculations. 

We use the following conventions for the Fourier transform. For a function $f:\mathbb{R}^d \to \mathbb{R}$, the Fourier transform and its inverse are defined by
\begin{align}
F(u) = F[f(x)](u) &= \int_{\mathbb{R}^d} e^{-i u^{\top} x} f(x)\,dx, \label{FT_def}\\[6pt]
f(x) = F^{-1}[F(u)](x) &= \frac{1}{(2\pi)^d} \int_{\mathbb{R}^d} e^{i x^{\top} u} F(u)\,du . \label{IFT_def}
\end{align}
We also use the {\bf convolution theorem}, which states that for functions
$f,g : \mathbb{R}^d \to \mathbb{R}$,
\begin{equation}\label{conv_theorem}
F[f * g](u) = F[f](u)\, F[g](u),
\end{equation}
where
\begin{equation}
(f * g)(x) = \int_{\mathbb{R}^d} f(x-y) g(y)\, dy.
\end{equation}
This identity underlies the convolution method developed in this paper, since the option value can be written as the convolution of the payoff function with the transition density of the log-price process, following the formulation introduced in \citet{hyndman2017convolution}.

We consider two implementations of the convolution method. In the first, which we refer to as CFFT-I, the convolution is applied directly to the option function using the characteristic function of the log-price increment. In the second, CFFT-II, we use the two-dimensional characteristic function of $(X_t,V_t)$ and integrate out the variance analytically. Both methods use the FFT to evaluate the convolution efficiently, but CFFT-II can offer improved stability for certain parameter values.

The premise of the convolution method is that the conditional probability density $\phi(x_t \mid x,v)$ depends only on the increment $x_t - x$,
\begin{equation}\label{premisze_condition}
    \phi(x_t \mid x,v) = \phi(x_t - x \mid v),
\end{equation}
as an approximation valid for short time increments $\Delta t$, consistent with the locally Gaussian behavior of the Heston transition density.
Although the notation in \eqref{premisze_condition} resembles the translation-invariance assumption used in the CONV method of \citet{lord2008fast}, the underlying meaning is different. In their setting, the relation $f(y \mid x) = f(y-x)$ arises from a L\'{e}vy-process representation and holds globally in maturity, leading to a convolution in the strike variable within the Carr and Madan framework. In our setting, \eqref{premisze_condition} is understood in the short-time Gaussian limit of the Heston model and is used to express the option value as an analytical convolution of the payoff with the transition density. This density-based formulation differs fundamentally from the strike-space convolution in \citet{lord2008fast} and plays a distinct role in the development of the CFFT methods introduced below.

In the Heston model, the short-time behavior of the transition density can be studied using the Fokker--Planck equation (see \citet{risken1996fokker}). As shown by \citet{dragulescu2002probability}, over a small time step $\Delta t$ the conditional distribution of $x_t$ is well approximated by a Gaussian with variance $v$:
\begin{equation*}
    \phi(x_t \mid x,v)
    = \frac{1}{\sqrt{2\pi v \Delta t}}
      \exp\!\left(
        -\frac{(x_t - x - (r - \tfrac12 v)\Delta t)^2}{2v\Delta t}
      \right)
    = \phi(x_t - x \mid v).
\end{equation*}
This convolution structure holds exactly under the short-time Gaussian approximation and numerically provides the basis for evaluating the conditional probabilities $P_i$.

	Applying the Fourier transform to $P_i$ and using the convolution theorem on the target function and the density function, we have 
        \begin{align} 
		F\left[P_i(x)\right](p) =& F\left[\mathbb{E}_i \left[\mathbf{1}_{S_T\geq K}\left|x=\ln{({S}/{K})}\right.\right]\right](p)
		= F\left[\int_\mathbb{R}\delta(y) \phi_i(y|x) dy \right](p) \nonumber\\
		=& F\left[\left(\delta(y) * \phi_i(y-x)\right)(x)\right](p)\nonumber\\ 
		=& F\left[\left(\delta(y) * \phi_i(-y)\right)(x)\right](p) \nonumber \\
		=& F\left[\delta(y)\right](p)F\left[\phi_i(-y)\right](p),\label{conv_result} 
	\end{align} 
	where the $\delta(\cdot)$ denotes the indicator function 
	\begin{equation*} 
	  \delta(x) =
                \begin{cases}
			1, &\text{if } \ x\geq0\\ 
			0, &\text{otherwise}. 
                \end{cases}
	\end{equation*} 
	The Fourier transform of the density function in (\ref{conv_result}) is 
	\begin{align} 
		F\left[\phi_i(-y)\right](p) =& \int_\mathbb{R} e^{-\iu p y} \phi(-y) dy 
		= \int_\mathbb{R} e^{\iu p (y-x)} \phi_i(y-x) dy\nonumber\\ 
		=& e^{-\iu p x}\int_\mathbb{R} e^{\iu p y} \phi_i(y\left|x\right.) dy
		= e^{-\iu p x}\mathbb{E}_i\left[e^{\iu p x_T}\left|x\right.\right]\nonumber\\ 
		=& e^{-\iu p x}\varphi_i(p)
		= \psi_i(p).\label{kernel} 
	\end{align} 
	We simplify (\ref{conv_result}) as 
	\begin{equation*} 
		F\left[P_i(x)\right](p) = F\left[\delta(x)\right](p)\psi_i(p), 
	\end{equation*} 
	and recover $P_i$ by 
	\begin{equation}\label{Pi_Fourier} 
		P_i(x) =  F^{-1}\left[F\left[\delta(x)\right](p)\psi_i(p)\right]. 
	\end{equation} 
	We apply the change of variables to $x=\ln\frac{S}{K}$ with varying $S$ and obtain the pricing formula to (\ref{Heston_call}) by the discrete Fourier transform:
	\begin{align} 
		C(S, K, v, t) =& S P_1\left(S,K\right) - Ke^{-r\tau}P_2\left(S,K\right)\nonumber\\ 
		=&S F^{-1}\left[F\left[\delta(x)\right](p)\psi_1(p)\right](x) - Ke^{-r\tau}F^{-1}\left[F\left[\delta(x)\right](p)\psi_2(p)\right](x)\nonumber\\ 
		\approx& S \tilde P_1 - Ke^{-r\tau} \tilde P_2,\label{tilde_C} 
	\end{align} 
	where the discretization of the real space is 
	\begin{equation*} 
		x_n = \left(n-\frac{N}{2}\right)\Delta x, \text{ for } n=0,1,\cdots,N-1, \text{ and } \Delta x = \frac{L}{N}, 
	\end{equation*} 
	and the discretization of the frequency space is 
	\begin{equation*} 
		p_n = \left(n-\frac{N}{2}\right)\Delta p, \text{ for } n=0,1,\cdots,N-1, \text{ and } \Delta p = \frac{2\pi}{L}. 
	\end{equation*}
        For a grid function $f(x_n, y_m)$ defined on an $N \times M$ uniform lattice, 
the discrete Fourier transform (DFT) and its inverse are given by
\begin{align}
\mathcal{D}[f](u_i, v_j) &= 
\sum_{k=0}^{N-1} \sum_{l=0}^{M-1}
    e^{-i\left(\frac{ki}{N} + \frac{lj}{M}\right)} f(x_k, y_l), 
\label{eq:DFT} \\
\mathcal{D}^{-1}[F](x_k, y_l) &= 
\frac{1}{NM}
\sum_{i=0}^{N-1} \sum_{j=0}^{M-1}
    e^{i\left(\frac{ki}{N} + \frac{lj}{M}\right)} F(u_i, v_j).
\label{eq:iDFT}
\end{align}

	The CFFT estimation of $\tilde P_i$ using the formula given in equation (\ref{Pi_Fourier}) is given by 
	\begin{equation}\label{tilde_P} 
		\tilde P_i = (-1)^{n} \mathcal{D}^{-1}\left[\left\{w_k\mathcal{D}\left[\left\{w_n(-1)^{n}\delta(x_n)\right\}_{n=0}^{N-1}\right](p_{k})\psi_i\left(p_{k}\right)\right\}_{k=0}^{N-1}\right]_{n}, 
	\end{equation}
        where $w_n$ denotes the standard trapezoidal weights on the interval $[-L/2,L/2]$.  Here $\mathcal{D}$ and $\mathcal{D}^{-1}$ in  to represent application of these discrete transforms to the spatial grid.
	\begin{remark} 
		The option price given by (\ref{tilde_C}) is similar to the Black-Scholes model except that the probability terms $\tilde P_1$ and $\tilde P_2$ do not have explicit formulas. In our numerical approach, we use (\ref{tilde_P}) to estimate the value of the probability terms. We denote this approach as the CFFT-I method. 
	\end{remark} 
 
Although the convolution method is efficient, the use of a truncated spatial domain introduces boundary error when the payoff is unbounded or nonperiodic on $[-L/2, L/2]$. This typically appears as oscillations near the boundaries of the interval and can degrade the accuracy of FFT-based convolution. To mitigate this effect, we adopt the damping and shifting transformations introduced in \cite{hyndman2017convolution} and adapt them to the Heston model. These transformations reduce boundary error and improve the numerical stability of both CFFT-I and CFFT-II.

Next, we introduce the CFFT-II method. To make the target function bounded and integrable, we introduce a damping parameter and write the Fourier transform of the convolution as a product
\begin{align}\label{Conv_method}
  F\left[e^{\alpha x} C(x)\right](p)
  =& e^{-r\tau} \int_{\mathbb{R}} e^{-ipx} e^{\alpha x}
  \, \mathbb{E}^{\mathbb{Q}}\!\left[(Ke^{x_T}-K)^+ \mid x\right] dx \nonumber \\
  =& e^{-r\tau} \int_\mathbb{R} e^{-\iu p x} e^{\alpha x}
      \int_\mathbb{R} g(y) \phi_2(y - x)\,dy\,dx \nonumber\\
  =& e^{-r\tau}\,F\left[e^{\alpha x}g(x)\right](p)\,\psi_2(p+\alpha\iu),
\end{align}
where $g(x) = \left(e^x - K\right)^+$.

The call option pricing function is obtained by inverting and undamping \eqref{Conv_method},
\begin{equation}\label{scheme2}
  C(x) = e^{-r\tau-\alpha x}\,
  F^{-1}\left[F\left[e^{\alpha x}g(x)\right]\psi_2(p+\alpha\iu)\right](x).
\end{equation}
We denote the approach based on equation~\eqref{scheme2} as the CFFT-II method. In the next sections, we first present the error analysis and then introduce two methods to improve the boundary error of the CFFT method.

	\subsection{Error analysis} 
	Let $\tilde{C}  = S \tilde P_1 - Ke^{-r\tau} \tilde P_2$ denote the convolution based approximation to the call option, with remining life $\tau$, using the convolution result in (\ref{tilde_C}). 
To analyze the associated truncation and discretization errors, we first examine the Fourier series expansion and the decay properties of the characteristic function.  Firstly, we investigate the Fourier expansion of a piece-wise smooth function $f$ with finite limiting point on $[-\frac{L}{2},\frac{L}{2}]$ 
	\begin{equation}\label{F_seires} 
		f(x) = \sum_{j=-\infty}^{\infty} F_j e^{-\iu j \frac{2\pi x}{L}}, 
	\end{equation} 
	with the coefficients $F_j$ 
	\begin{equation*} 
		F_j = \frac{1}{L}\int_{-\frac{L}{2}}^{\frac{L}{2}} f(x) e^{\iu j \frac{2\pi x}{L}} dx. 
	\end{equation*} 
	The Fourier coefficients $F_j\rightarrow 0$ as $|j|\rightarrow\pm\infty$ and  $|F_j|\leq \bar{f}$ 
        when $f$ is bounded on $[-\frac{L}{2},\frac{L}{2}]$. 
        Thus we can bound the modulus of $F_j$ by 
	\begin{equation}\label{F_boundary} 
		\left|F_j\right| \leq \min\left(\bar f, \frac{\epsilon(L)}{\left|j\right|}\right), 
	\end{equation} 
	for a positive bounding constant $\epsilon(L)$,  depending only on $L$. 
	 
	Usually, the characteristic function of the Black-Scholes model decays as $\exp\left(-cx^2\right)$ and that of the Heston model has exponential decays as $\exp(-c|x|)$ for some constant value of $c$ as discussed in \citet{lord2007optimal}. We summarize the asymptotic behavior of the characteristic function (\ref{my_kernel}) in the following proposition. 
\begin{proposition}[Asymptotic characteristic function]\label{Prop_limiting_charac}
Assume that $\kappa$, $\theta$, $\sigma$, $v$, and $\tau$ are positive and that $\rho\in(-1,1)$. Then the kernel function~\eqref{kernel} satisfies the asymptotic relation
\[
\psi_i(p) \approx A_\infty\, e^{\iu B_\infty}\, \exp\!\left(-D|p|\right), \qquad p \to \infty,
\]
where
\[
A_\infty = \left(4(1-\rho^2)\right)^{a/\sigma^2}, \qquad
B_\infty = \frac{2a}{\sigma^2} \arcsin\!\left(\sgn(p)\rho\right)
           - \frac{\rho}{\sigma}\left(v + \sgn(p)a\tau\right)p, \qquad
D = \frac{\sqrt{1-\rho^2}}{\sigma}\left(v+a\tau\right) > 0.
\]
\end{proposition}
	\noindent {\it The proof is given in Appendix \ref{Pf_prop_limiting_charac}.} 

By Proposition~\ref{Prop_limiting_charac}, we can bound the modulus of the characteristic function by
\begin{equation}\label{psi_bound}
   |\psi_i(p)| \le \epsilon A_\infty e^{-D|p|},
\end{equation}
for some positive constant $\epsilon$. The exponential decay of $|\psi_i(p)|$ implies that the Fourier integrand becomes negligible outside a finite truncation interval, which justifies the CFFT truncation. Using \eqref{psi_bound} together with the boundary estimate \eqref{F_boundary}, we next derive bounds for both the truncation error and the discretization error. The following theorem provides an error bound for $|C - \tilde C|$.
\begin{theorem}[Error of the convolution method]\label{thm_error_bound}
Let $f$ be an integrable function that is bounded by $\bar f$ on 
$[-\tfrac{L}{2}, \tfrac{L}{2}]$. Under the measure $P_i$, the error between the true value
\[
E(x) = \mathbb{E}^{P_i}\!\left[f(x_T)\,\middle|\, x_0 = x\right],
\]
and its CFFT approximation
\[
\tilde E(x_n)
 = (-1)^{n}\,
   \mathcal{D}^{-1}
   \left[
      \left\{
         w_k\,
         \mathcal{D}
         \left[
            \left\{ w_n (-1)^{n} f(x_n) \right\}_{n=0}^{N-1}
         \right](u_k)
         \psi_i(p_k)
      \right\}_{k=0}^{N-1}
   \right]_{n},
\]
on the truncation interval $[-\tfrac{L}{2}, \tfrac{L}{2}]$ with discretization parameters
\[
x_n = \left(n - \tfrac{N}{2}\right)\Delta x,\quad n = 0,\dots,N-1,\qquad
\Delta x = \tfrac{L}{N},
\]
\[
p_n = \left(n - \tfrac{N}{2}\right)\Delta p,\quad n = 0,\dots,N-1,\qquad
\Delta p = \tfrac{2\pi}{L},
\]
is bounded by
\[
|E - \tilde E|
   \le \epsilon_1\, e^{-\frac{\pi D}{L}N}
     + \epsilon_2\, N^{-m},
\]
where
\[
\epsilon_1 = \frac{L A_\infty \bar f\, e^{\frac{2\pi D}{L}}}{\pi D}\,\epsilon_{v,\tau},
\qquad
\epsilon_2 = \frac{L A_\infty}{\pi D}\,\epsilon_L\, \epsilon_{v,\tau},
\]
for some positive constants $\epsilon_{v,\tau}$ and $\epsilon_L$.
\end{theorem}
	\noindent {\it The proof is given in Appendix~\ref{Pf_thm_error_bound}.} 
\noindent

Applying Theorem~\ref{thm_error_bound}, we obtain the following bound for 
$|P_i - \tilde P_i|$:
\[
|e_i|
 = |P_i - \tilde P_i|
 \le \frac{L A_\infty \bar f\, e^{\frac{2\pi D}{L}}}{\pi D}\,\epsilon_{v,\tau}\,
       e^{-\frac{\pi D}{L}N}
   + \frac{L A_\infty}{\pi D}\,\epsilon_L\,\epsilon_{v,\tau}\, N^{-m}.
\]
Using this estimate for $i=1,2$, we obtain an error bound for the pricing error of a call option in the Heston model under the CFFT-I method:
\begin{align}\label{error_c}
|e(x)| 
 &= |C(x) - \tilde C(x)| \nonumber\\
 &= \left| K e^{x}\left(P_1(x) - \tilde P_1(x)\right)
      - K e^{-r\tau}\left(P_2(x) - \tilde P_2(x)\right) \right| \nonumber\\
 &\le K e^{x}\, |e_1| + K e^{-r\tau}\, |e_2| \nonumber\\
 &\le K\left(e^{x} + e^{-r\tau}\right)
     \left(
        \frac{L A_\infty \bar f\, e^{\frac{2\pi D}{L}}}{\pi D}\,\epsilon_{v,\tau}\,
        e^{-\frac{\pi D}{L}N}
        + \frac{L A_\infty}{\pi D}\,\epsilon_L\,\epsilon_{v,\tau}\, N^{-m}
     \right).
\end{align}
A similar bound can be derived for the CFFT-II method. We summarize the resulting error estimate for $|e|$ in the following corollary.

\begin{corollary}
For the Heston call option, the CFFT-I and CFFT-II methods satisfy the error estimate
\[
|e| \le \mathcal{O}\!\left(e^{-\frac{\pi D}{L} N}\right) 
       + \mathcal{O}\!\left(N^{-m}\right),
\]
for any $m \ge 2$.
\end{corollary}

\begin{note}
We observe that the discretization error is at least second order, consistent with the findings of \citet{lord2008fast}, while the truncation error decays exponentially with the frequency. Equation~\eqref{error_c} also indicates that the boundary errors increase as $x$ approaches $\pm L/2$, which motivates the boundary-control schemes introduced in the next subsection.
\end{note}

Unlike the Carr and Madan and COS methods, for which truncation errors are typically assessed empirically, we obtain an explicit analytical error bound for the convolution approximation. The bound shows exponential decay of the truncation error and second-order convergence of the discretization error, providing a rigorous theoretical foundation for the proposed CFFT method.

	\subsection{Boundary control for the CFFT method} 
        Our first consideration when applying the Fourier transform to option pricing is the feasibility of the transform.
        Sufficient conditions for successfully applying Fourier transform require the target function to be $L_1$-integrable.
        However, the call option payoff is not $L_1$-integrable with respect to either the log-price or the log-strike.
        Nevertheless, we can still apply the Fourier transform to the target function on a truncated region by introducing a damping parameter, which attenuates the nonintegrable tail toward zero.   Equation (\ref{scheme2}) is well-defined provided that the $(\alpha + 1)^{\text{th}}$ moment of $S_{t}$ exists, as pointed out by \citet{lord2007optimal} 
	\begin{equation*} 
		\left|\varphi(p-(\alpha+1)\iu)\right| \leq \varphi(-(\alpha+1)\iu) = \mathbb{E}\left[S_T^{(\alpha+1)}\right]\le \infty. 
	\end{equation*}

	After ensuring feasibility, we next consider periodicity effects introduced by the Fourier transform.  Our second consideration concerns the periodicity effects introduced by the Fourier transform, since the inverse transform may distort a nonperiodic target function. Comparing the truncation on logarithms of the strike price in \citet{lord2007optimal}, our truncation on the logarithm of the stock price could lead to large boundary errors when the option is exponentially increasing as the underlying asset moves to deep in-the-money. Such problems can be found in \citet{hyndman2017convolution}, however, they introduced a shifting method on the target function to address the boundary error. The basic idea of shifting the target function is to map it from non-periodic to a periodic function which would be considered as a real signal. The shifting method requires a function $h(x)$ with explicit expectation $\mathbb{E}\left[h(x_t)\left|x\right.\right]$. Thus the candidate for shifting function $h(x)$ can be chosen from polynomial and exponential functions. \citet{hyndman2017convolution} suggest the first order polynomial as the shifting function $h(x)=Ax + B$ such that the damping of the shifted target function $\tilde f^\alpha(x) = e^{\alpha x} \left(f(x) - h(x)\right)$ is smoothly connected at the boundaries 
	\begin{align*} 
		\tilde f^\alpha(x_0) =& \tilde f^\alpha(x_n),\\ 
		\frac{d \tilde f^\alpha}{dx}(x_0) =& \frac{d \tilde f^\alpha}{dx}(x_n). 
	\end{align*}
In our implementation, shifting the call option by a linear function generates a kink at the money and does not perform well. We therefore propose an exponential shift function $h_{2}=Ae^{x}+B$ to ensure smooth damping near the boundaries. In CFFT-I we choose a linear function $h_1=Ax+B$ to shift the $\delta$ function and in CFFT-II we choose an exponential function $h_2=Ae^x + B$ to shift the call option which can also be applied in BSDE-based numerical methods. For CFFT-I, for $\alpha=0$, we have a {\bf linear-shift scheme}: 
\begin{equation}\label{shift1}
  h(x) = Ax+B, \qquad \tilde f^\alpha(x)= f(x) - h(x),
\end{equation}
where
\begin{equation*}
   A = \frac{f(x_N)-f(x_0)}{x_N - x_0}, \qquad B = \frac{x_N f(x_0) - x_0 f(x_N)}{x_N - x_0}.
\end{equation*}
For CFFT-II, for $\alpha<-1$, we have an {\bf exponential-shift scheme:} 
	\begin{equation}\label{shift2} 
	  h(x) = Ae^{x}+B, \qquad \tilde f^\alpha(x)= e^{\alpha x}\left(f(x) - h(x)\right),
          \end{equation}
        where
\begin{align*}
        f'_0 =& \frac{-3f(x_0) + 4f(x_1) - f(x_2)}{2\Delta x},	\qquad f'_N = \frac{3f(x_N) - 4f(x_{N-1}) + f(x_{N-2})}{2\Delta x},\\ 
	A    =& \frac{e^{\alpha x_N} f'_N - e^{\alpha x_0} f'_0}{e^{(\alpha+1) x_N} - e^{(\alpha+1) x_0}}, \qquad \text{ and } \qquad
			B = \frac{x_N f(x_0) - x_0 f(x_N)}{x_N - x_0}.
\end{align*}
        These transformations ensure continuity and smoothness of the damped target function at domain boundaries, significantly improving numerical stability. 
	 
	We can recover CFFT-I by reversing the shifting scheme (\ref{shift1}) 
	\begin{align*} 
		\mathbb{E}^{\mathbb{P}_i}\left[f(x_T)\left|x\right.\right] =& \mathbb{E}^{\mathbb{P}_i}\left[\tilde f(x_T)\left|x\right.\right] + \mathbb{E}^{\mathbb{P}_i}\left[h(x_T)\left|x\right.\right]\\ 
		=& F^{-1}\left[F\left[\tilde f(x)\right](p)\psi_i(p)\right](x) + A\mathbb{E}\left[x_T\left|x\right.\right] + B\\ 
		=& F^{-1}\left[F\left[\tilde f(x)\right](p)\psi_i(p)\right](x) - \iu A \frac{\partial\varphi_i}{\partial p}(0) + B, 
	\end{align*} 
	and recover CFFT-II by reversing the shifting scheme (\ref{shift2}) 
	\begin{align*} 
		\mathbb{E}^{\mathbb{P}_2}\left[f(x_T)\left|x\right.\right] =& \mathbb{E}^{\mathbb{P}_2}\left[\tilde f(x_T)\left|x\right.\right] + \mathbb{E}^{\mathbb{P}_2}\left[h(x_T)\left|x\right.\right] 
		= e^{-\alpha x}F^{-1}\left[F\left[e^{\alpha x}\tilde f(x)\right](p)\psi_2(p)\right](x) + A\mathbb{E}\left[e^{x_T}\left|x\right.\right] + B\\ 
		=& e^{-\alpha x}F^{-1}\left[F\left[e^{\alpha x}\tilde f(x)\right](p)\psi_2(p)\right](x) + A\varphi_2(-\iu) + B. 
	\end{align*} 
	
	In the next section, we present the numerical results of the CFFT-I and CFFT-II methods applied to pricing problems in the Heston model.

	\section{Numerical results}\label{sec3} 
        We assess the accuracy of the proposed methods by comparing numerical results to the semi-closed-form Heston solution (\ref{integration}), and evaluate computational efficiency relative to the Carr and Madan FFT method. We first present the results of the CFFT-I method applied to the estimation of the probabilities in the Heston model. We then apply CFFT-II to price the European call option and illustrate the effect of the boundary control schemes.
We then summarize the performance of CFFT-II across a set of representative strikes. The comparison between the CFFT and FFT methods is conducted on both the log-stock and log-strike domains.

The truncated spatial domain is centered at $\log(S/K)$, with $N$ grid points
on the interval $[x_0, x_N] = \log(S/K) + [-L/2, L/2]$.  Similarly, the interval  $[x_0,x_N]=\log(K/S) + [-L/2, L/2]$ is used for the log-strike domain.  The parameters used are $r=0.03$, $v = 0.1$, $\Lambda = 1$, $\rho = -0.8$, $\kappa = 3$, $\theta = 0.1$, $\sigma = 0.25$, $T=1$, $L=10$, and $N=2000$. 
	\begin{figure}[ht] 
		\centering 
		\hspace*{-0cm} 
		\caption{Comparison of CFFT-I probabilities $P_1$ and $P_2$ with and without the shifting scheme.} 
		\includegraphics[width=0.75\linewidth]{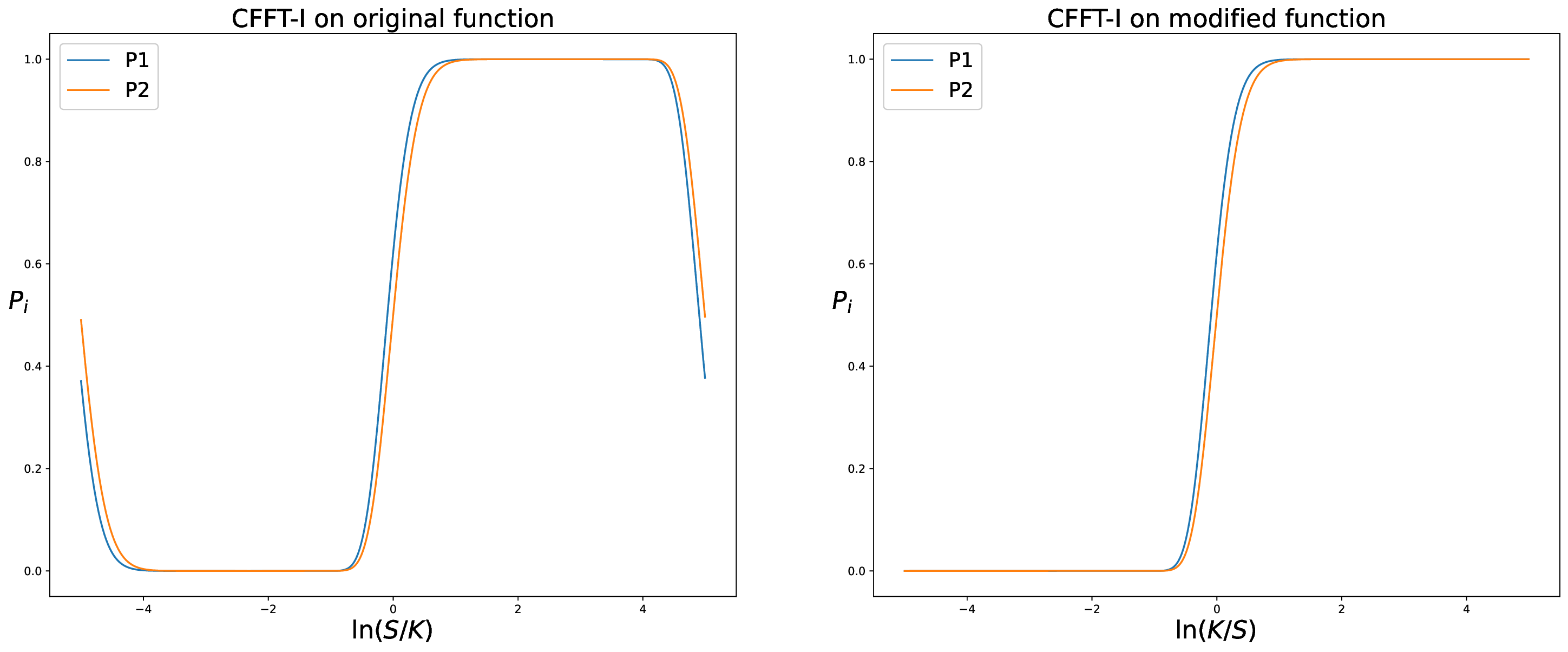} \label{chap2:pic2}
	\end{figure} 
Figure~\ref{chap2:pic2} illustrates the effect of the shifting scheme. The left panel shows the raw CFFT-I probabilities, and the right panel shows the results with shifting applied. The shifting scheme eliminates boundary oscillations near the truncation endpoints and improves numerical stability.  We observe that the boundary values at $x_0$ and $x_N$ are accurately controlled when the shifting scheme is used:  $P_1(x_0) = 0$, $P_2(x_0) = 0$, $P_1(x_N) = 0.99999844$, and $P_2(x_N) = 0.99999839$. 
	Inclusion of the shifting scheme clearly improves numerical stability and accuracy at the boundaries.

\begin{figure}[ht] 
		\centering 
		\hspace*{-0cm}  
		\caption{Error of CFFT-I} 
		\includegraphics[width=0.75\linewidth]{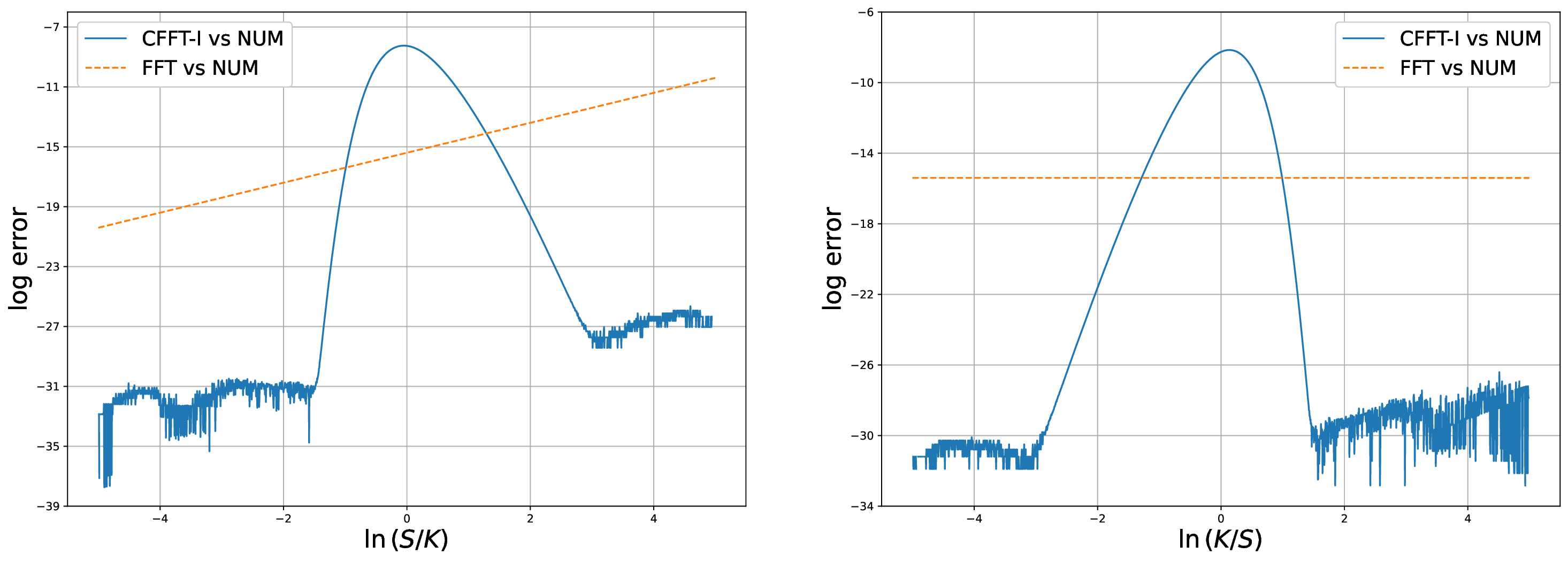} 
		\caption*{$r=0.03$, $v = 0.1$, $\Lambda = 1$, $\rho = -0.8$, $\kappa = 3$, $\theta = 0.1$, $\sigma = 0.25$, $T=1$, $L=10$, $N=2000$} 
		\label{chap2:pic3} %
		\vspace*{-0.25cm}  
	\end{figure}

        Figure \ref{chap2:pic3} examines the accuracy of the CFFT-I method by comparing the results with both the FFT method and a numerical integration benchmark (denoted as ``NUM''). The solid lines represent the error between the CFFT-I method and the numerical integration method, and the dashed lines represent the error between the FFT method and the numerical integration method. The CFFT-I method outperforms the FFT method except at the money, where the FFT method retains slightly higher precision.        
	 
	\begin{figure}[ht] 
		\centering 
		\vspace*{-0cm}  
		\caption{Error of CFFT-II} 
		\includegraphics[width=0.75\linewidth]{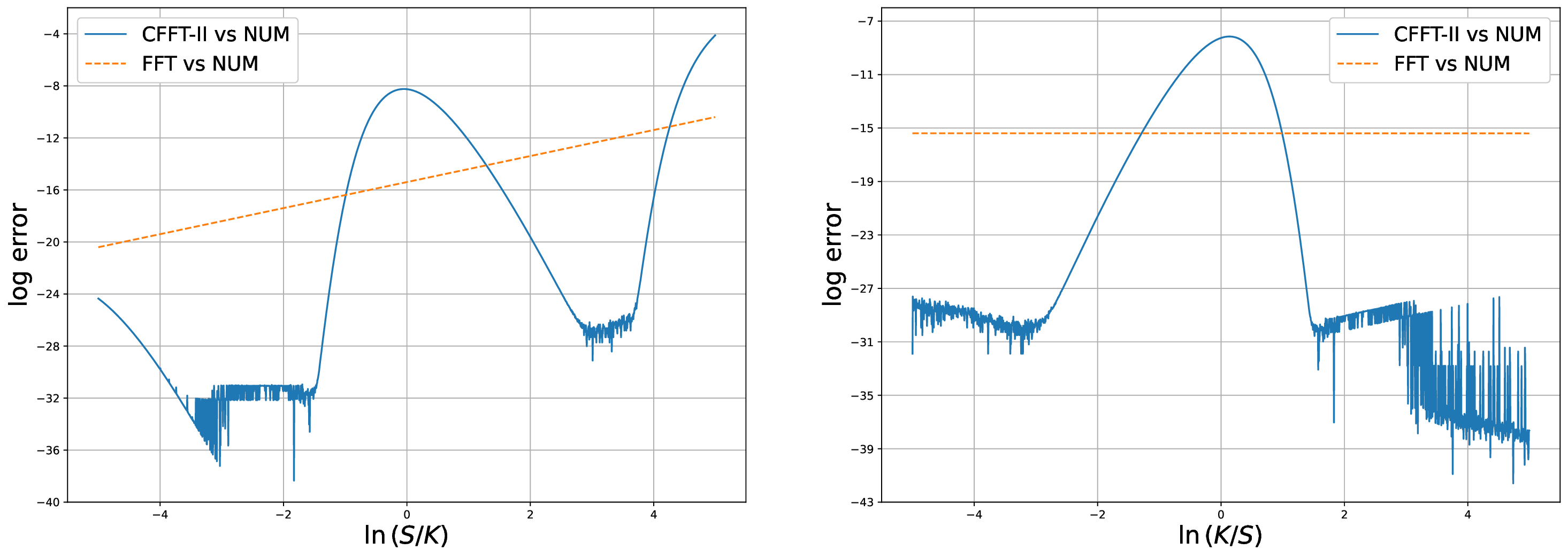} 
		\caption*{$r=0.03$, $v = 0.1$, $\Lambda = 1$, $\rho = -0.8$, $\kappa = 3$, $\theta = 0.1$, $\sigma = 0.25$, $T=1$, $L=10$, $N=2000$, $\alpha=-2$} 
		\label{chap2:pic4} %
		\vspace*{-0.25cm}  
	\end{figure} 
 
	\begin{figure}[ht] 
		\centering 
		\hspace*{-0cm}  
		\caption{CFFT-II error with different damping and shifting schemes.} 
		\includegraphics[width=0.75\linewidth]{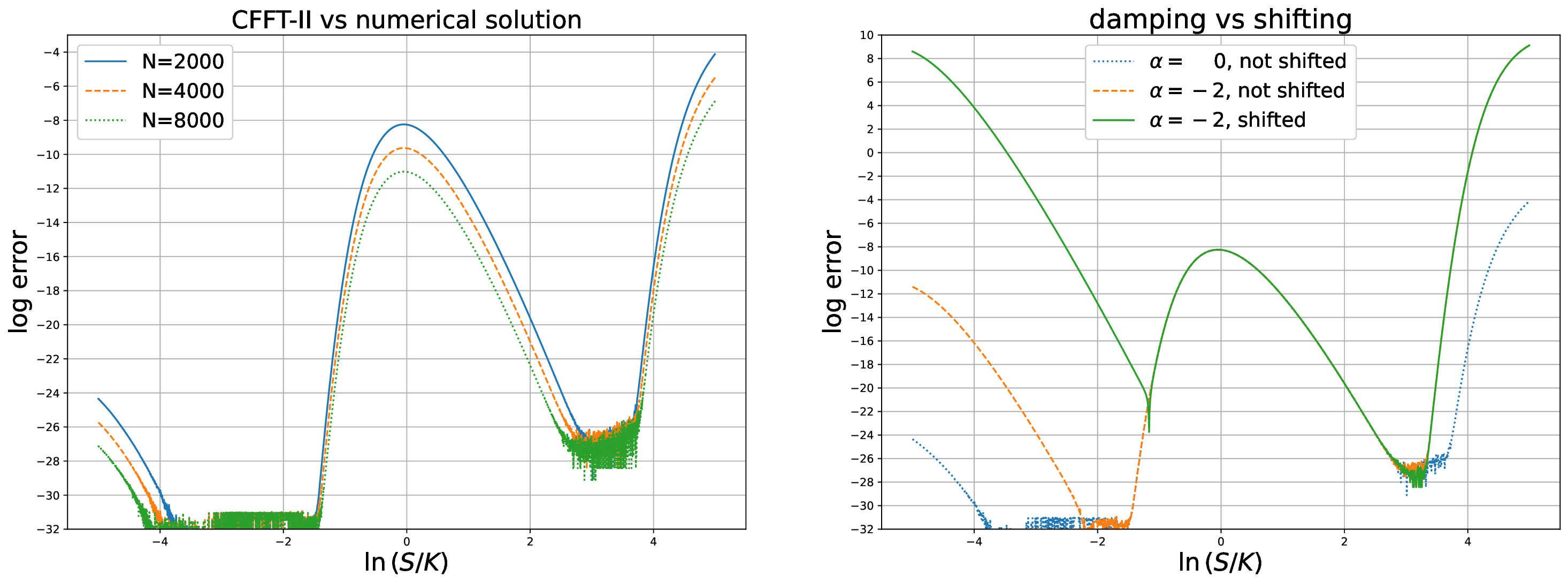} 
		\caption*{$r=0.03$, $v = 0.1$, $\Lambda = 1$, $\rho = -0.8$, $\kappa = 3$, $\theta = 0.1$, $\sigma = 0.25$, $T=1$, $L=10$} 
		\label{chap2:pic5} %
		\vspace*{-0.25cm}  
	\end{figure} 
	 
	\begin{table}[ht] 
		\centering 
		\caption{CPU time, call option values, and absolute errors relative to the semi-closed-form Heston solution for strikes $K=80, 100, 120$.}
                \label{chap2:Table1} 
		\vspace*{-0.25cm} 
		\begin{adjustbox}{width=0.75\columnwidth,center} 
			\begin{tabular}{cccccccccc} 
				\hline 
				&\multicolumn{2}{|c}{CPU time (ms)}&\multicolumn{2}{|c}{S=100,K=80}&\multicolumn{2}{|c}{S=100,K=100}&\multicolumn{2}{|c}{S=100,K=120}\\ 
				\hline 
				&\multicolumn{1}{|c}{CFFT-II}&\multicolumn{1}{|c}{FFT}&\multicolumn{1}{|c}{call}&\multicolumn{1}{|c}{error}&\multicolumn{1}{|c}{call}&\multicolumn{1}{|c}{error}&\multicolumn{1}{|c}{call}&\multicolumn{1}{|c}{error}\\ 
				\hline 
				N=2000 & 0.124 & 0.155 & 25.77846 & 5.93e-05 & 13.45867 & 2.60e-04 & 5.97903 & 1.40e-04 \\ 
				N=4000 & 0.175 & 0.294 & 25.77841 & 8.04E-06 & 13.45887 & 6.50e-05 & 5.97885 & 4.29e-05 \\ 
				N=8000 & 0.251 & 0.544 & 25.77841 & 4.60e-06 & 13.45892 & 1.63e-05 & 5.97889 & 4.73e-06 \\ 
				\hline 
			\end{tabular} 
		\end{adjustbox} 
		\caption*{$r=0.03$, $v = 0.1$, $\Lambda = 1$, $\rho = -0.8$, $\kappa = 3$, $\theta = 0.1$, $\sigma = 0.25$, $T=1$, $L=10$, $\alpha=-2$} 
	\end{table}

The CFFT-II method achieves comparable accuracy to the FFT benchmark with less than half the computational cost. The convergence behavior observed across increasing N aligns with the theoretical error analysis from Section 3.1.  Similar to Figure \ref{chap2:pic3}, Figure \ref{chap2:pic4} shows the log-error for the CFFT-II method.  The CFFT-II method is faster than CFFT-I, but boundary errors are more pronounced, particularly when the shifting scheme is omitted. The left panel of Figure  \ref{chap2:pic5} shows that the accuracy increases as the discretization $N$ increases, and that the boundary error is controlled by the damping and shifting schemes, consistent with the error analysis.  The right panel of Figure~\ref{chap2:pic5}  illustrates the effect of combining damping and shifting. Without these schemes, the error grows rapidly near both boundaries, while the combined approach yields stable and accurate results across the domain. The comparison among different choices of damping and shifting parameters in Figure \ref{chap2:pic5} indicates that the damping parameter primarily reduces the left-boundary error, while the shifting parameter reduces the right-boundary error when the Fourier transform is applied to an unbounded and nonperiodic function.

Table  \ref{chap2:Table1} presents numerical results for the CFFT-II and FFT methods on the log-stock domain with different values of $N$. The  option values for strikes $K = 80$, $K = 100$, and $K = 120$ are $25.77840$, $13.45893$, and $5.97889$, respectively. The FFT method, which behaves like an enhanced numerical integration method, is relatively insensitive to $N$, and the errors remain on the order of $2.06 \times 10^{-7}$ for all cases. In contrast, the CFFT-II method converges as $N$ increases and requires less computation time. The CPU times in Table  \ref{chap2:Table1} indicate that CFFT-II has a significant computational speed advantage over FFT.

Numerically, the CFFT method achieves comparable or superior accuracy with smaller grids and reduced computation time, while eliminating the need for damping-parameter calibration required in the Carr and Madan approach. Furthermore, unlike the COS method, the CFFT method handles discontinuous or non-smooth payoffs without oscillatory artifacts near the truncation boundaries.

\section{Conclusion}\label{sec4}
In this paper we developed a convolution–FFT method for option valuation under the Heston model. The key ingredients are a continuously differentiable representation of the characteristic function and an efficient convolution formulation on a truncated spatial domain. This approach eliminates the discontinuity issues of the classical Heston characteristic function and improves numerical stability in Fourier-based valuation. The availability of closed-form truncation and discretization error bounds further distinguishes the method from previous Heston FFT formulations, which rely primarily on empirical or heuristic error assessment.

We presented two implementations, CFFT-I and CFFT-II, and analyzed the boundary errors introduced by truncation. A combination of damping and shifting transformations was introduced to reduce these errors, with numerical experiments demonstrating that both methods provide accurate valuations and that the CFFT-II variant is particularly efficient for large-scale computations.

The convolution–FFT framework therefore offers a stable and mathematically justified alternative to existing Fourier-based pricing methods for the Heston model. By combining a smooth characteristic function with explicit analytical error bounds, the method provides both numerical reliability and practical guidance for parameter selection. These features make the convolution method a promising foundation for extensions to multi-factor volatility models, early-exercise features, and other complex derivative structures.

        \nocite{XGao-PHDthesis2021} 
	\bibliographystyle{abbrvnat} 
	\bibliography{newbib}

@phdthesis{XGao-PHDthesis2021,
      title = {Stochastic control, numerical methods, and machine learning in finance and insurance},
      author = {Gao, Xiang},
      month = {March},
      year = {2021},
      school = {Concordia University},
      url = {https://spectrum.library.concordia.ca/id/eprint/988412/},
}

@book{risken1996fokker,
	title={The {F}okker-{P}lanck Equation},
	subtitle = {Methods of Solution and Applications},
	author={Risken, Hannes},
	pages={63--95},
	year={1996},
	publisher={Springer},
	edition = {2nd},
}

@article{black1973pricing,
	title={The pricing of options and corporate liabilities},
	author={Black, Fischer and Scholes, Myron},
	journal={Journal of Political Economy},
	volume={81},
	number={3},
	pages={637--654},
	year={1973},
	publisher={The University of Chicago Press}
}

@article{carr1999option,
	title={Option valuation using the fast {F}ourier transform},
	author={Carr, Peter and Madan, Dilip},
	journal={Journal of Computational Finance},
	volume={2},
	number={4},
	pages={61--73},
	year={1999}
}

@book{fouque2000derivatives,
        title={Derivatives in Financial Markets with Stochastic Volatility},
        author={Fouque, Jean-Pierre and Papanicolaou, George and Sircar, K Ronnie},
        year={2000},
        publisher={Cambridge University Press},
}

@article{dragulescu2002probability,
	title={Probability distribution of returns in the {H}eston model with stochastic volatility},
	author={Dr\u{a}gulescu, Adrian A and Yakovenko, Victor M},
	journal={Quantitative Finance},
	volume={2},
	number={6},
	pages={443--453},
	year={2002},
}

@article{cui2017full,
	title={Full and fast calibration of the {H}eston stochastic volatility model},
	author={Cui, Yiran and del Ba{\~n}o Rollin, Sebastian and Germano, Guido},
	journal={European Journal of Operational Research},
	volume={263},
	number={2},
	pages={625--638},
	year={2017},
}

@article{feller1951two,
	title={Two singular diffusion problems},
	author={Feller, William},
	journal={Annals of Mathematics},
	pages={173--182},
	year={1951},
	publisher={JSTOR}
}

@article{heston1993closed,
	title={A closed-form solution for options with stochastic volatility with applications to bond and currency options},
	author={Heston, Steven L},
	journal={The Review of Financial Studies},
	volume={6},
	number={2},
	pages={327--343},
	year={1993},
	publisher={Oxford University Press}
}

@article{hyndman2017convolution,
	title={A convolution method for numerical solution of backward stochastic differential equations},
	author={Hyndman, Cody B and Oyono~Ngou, Polynice},
	journal={Methodology and Computing in Applied Probability},
	volume={19},
	number={1},
	pages={1--29},
	year={2017},
	publisher={Springer}
}

@article{lord2007optimal,
	title={Optimal {F}ourier Inversion in Semi-Analytical Option Pricing},
	author={Lord, Roger and Kahl, Christian},
	year = {2006},
	month = {01},
	pages = {},
	volume = {10},
	journal={Tinbergen Institute, Tinbergen Institute Discussion Papers}
}

@article{levendorskiui2012efficient,
	title={Efficient pricing and reliable calibration in the {H}eston model},
	author={Levendorski{\u\i}, Sergei},
	journal={International Journal of Theoretical and Applied Finance},
	volume={15},
	number={07},
	pages={1250050},
	year={2012},
}

@article{kahl2005not,
	title={Not-so-complex logarithms in the {H}eston model},
	author={Kahl, Christian and J{\"a}ckel, Peter},
	journal={Wilmott},
	volume={19},
	number={9},
	pages={94--103},
	year={2005}
}

@article{lord2010complex,
	title={Complex Logarithms in {H}eston-{L}ike models},
	author={Lord, Roger and Kahl, Christian},
	journal={Mathematical Finance},
	volume={20},
	number={4},
	pages={671--694},
	year={2010},
}

@incollection {MR3375192,
	AUTHOR = {Lucic, Vladimir},
	TITLE = {On singularities in the {H}eston model},
	BOOKTITLE = {Large Deviations and Asymptotic Methods in Finance},
	EDITORS = {Friz, Peter K and Gatheral, Jim and Gulisashvili, Archil and Jacquier, Antoine and Teichmann, Josef},
	SERIES = {Springer Proc. Math. Stat.},
	VOLUME = {110},
	PAGES = {439--448},
	PUBLISHER = {Springer},
	YEAR = {2015},
}

@article{lee2004option,
  author = {Lee, R.},
  title = {Option pricing by transform methods: Extensions, unification and error control},
  journal = {Journal of Computational Finance},
  volume = {7},
  number = {3},
  pages = {51--86},
  year = {2004}
}

@article{lord2008fast,
	title={A fast and accurate {FFT}-based method for pricing early-exercise options under {L}{\'e}vy processes},
	author={Lord, Roger and Fang, Fang and Bervoets, Frank and Oosterlee, Cornelis W},
	journal={SIAM Journal on Scientific Computing},
	volume={30},
	number={4},
	pages={1678--1705},
	year={2008},
}

@article{crocce2017error,
  author = {Crocce, Fabi\'n and H\"app\"ol\"a, Juho and Kiessling, Jonas and Tempone, Ra\'ul},
  title = {Error analysis in {F}ourier methods for option pricing},
  journal = {Journal of Computational Finance},
  volume = {21},
  number = {1},
  pages = {1--30},
  year = {2017},
  doi = {10.21314/JCF.2016.327}
}

	\appendix 
	 
	\section{Appendix} 
	In this appendix we provide the proofs of the technical results that would otherwise disrupt the flow of the paper.

	\subsection{Proof of Theorem \ref{thm_characteristic}} 
        \begin{proof}\label{Pf_theorem_thm_characteristic} 
We solve the PDE \eqref{characteristic2_PDE} by substituting the exponential affine expression introduced in Section~\ref{sec1} and matching coefficients of the terms in $p$ and $q$. This reduces the PDE to a system of ordinary differential equations for the functions appearing in the characteristic-function representation.   We may write 
		\begin{equation*} 
			\frac{\partial P_i}{\partial t} + \left(r + c_i v\right) \frac{\partial P_i}{\partial x} + (a - b_i v) \frac{\partial P_i}{\partial v} + \frac{1}{2}v \frac{\partial^2 P_i}{\partial x^2} + \frac{\sigma^2}{2} v \frac{\partial^2 P_i}{\partial v^2} + \rho\sigma v \frac{\partial^2 P_i}{\partial x \partial v} = 0, 
		\end{equation*} 
		with boundary conditions 
		\begin{equation*} 
			P_i(T, \boldsymbol{U}, \boldsymbol{X}) = e^{\iu \boldsymbol{U}^\top \boldsymbol{X}} = e^{\iu\left(px + qv\right)} 
		\end{equation*}
                for $i=1,2$ where $U=(p ,q)^\top$ and $X=(x, v)^\top$. 
	        We make an ansatz for $\psi(t, p, q, x, v)$ in the following form 
		\begin{equation*} 
			P_i(t, p, q, x, v) = \exp\left(A_i(t)x + B_i(t)v + C_i(t)\right), 
		\end{equation*} 
		where functions $A(t)$, $B(t)$ and $C(t)$ depend only on $t$ and satisfy boundary conditions $A(T) = \iu p$, $B(T) = \iu q$ and $C(T)=0$. 
		 
		Applying the ansatz to the PDE, we obtain the following ordinary differential equation:  
		\begin{equation}\label{ansatz2_PDE_2} 
			A_i'(t) x + \left( B_i'(t) + \frac{1}{2} A_i^2 + c_i A_i - b_i B_i + \frac{1}{2}\sigma^2 B_i^2 + \rho\sigma A_i B_i \right) v + C_i'(t) +  r  A_i(t) + a B_i(t) = 0. 
		\end{equation} 
		If equation (\ref{ansatz2_PDE_2}) holds for all $x\in\mathbb{R}$, $v\in(0,\infty)$ and $t\in[0,T]$, then we must have 
		\begin{align} 
			&A'(t) = 0,\label{ansatz2_alpha_ODE_2}\\ 
			&B'(t) + \frac{1}{2} A(t)^2  + c_i A(t) - b_i B(t) + \frac{\sigma^2}{2} B(t)^2 + \rho\sigma A(t) B(t) = 0,\label{ansatz2_beta_ODE_2}\\ 
			&C'(t) +  r A(t) + a B(t) = 0,\label{ansatz2_gamma_ODE_2} 
		\end{align} 
		so we obtain 
		\begin{equation}\label{ansatz2_alpha_2} 
			A(t) = \iu p. 
		\end{equation} 
		Substituting equation (\ref{ansatz2_alpha_2}) into equation (\ref{ansatz2_beta_ODE_2}) and simplifying, we obtain
		\begin{equation}\label{ansatz2_beta_ODE2_2} 
			B'(t) + \frac{1}{2}\sigma^2 B^2(t) - \left(b_i - \iu\sigma \rho p\right) B(t) - \frac{p^2 - 2\iu c_i p}{2} = 0. 
		\end{equation} 
		Equation (\ref{ansatz2_beta_ODE2_2}), with boundary value $B(T) = \iu q$, is Riccati equation with constant coefficients and has solution
		\begin{align}\label{ansatz2_beta_2} 
			B(t) =& \frac{\iu \gamma }{\sigma^2}\tan\left(\frac{\iu \gamma }{2}(T - t) + \vartheta\right) + \frac{b_i- \iu \sigma\rho p}{\sigma^2}, 
		\end{align} 
		where  $\vartheta =  \arctan\left({\iu\lambda}/{\gamma}\right)$, 
		       $\gamma  = \sqrt{\sigma^2\left(p^2 - 2\iu c_i p\right) + \left(b_i - \iu\sigma \rho p\right)^2}$, and
		       $\lambda = (b_i - \iu\sigma \rho p - \iu\sigma^2 q)$. 
		The solution of (\ref{ansatz2_gamma_ODE_2}) with boundary $C(T) = 0$ can be obtained by integration. With $\tau=T-t$, we obtain 
		\begin{align*} 
			C(t) =& \int_t^{T} \left(\iu p r  + a B(s)\right) ds 
			= \iu p r (T-t) + \frac{a\left(b_i - \iu p\rho\sigma\right)}{\sigma^2} (T - t) - \frac{2a}{\sigma^2} \ln \frac{\cos\left(\frac{\iu \gamma }{2}(T - t) + \vartheta\right)}{\cos\left( \vartheta \right)}\nonumber\\ 
			=& \iu p r \tau + \frac{a\left(b_i - \iu p\rho\sigma\right)}{\sigma^2}\tau - \frac{2a}{\sigma^2} \ln \frac{\cos\left(\frac{\iu \gamma\tau}{2}\right)\cos\left(\vartheta\right) - \sin\left(\frac{\iu \gamma\tau}{2}\right)\sin\left(\vartheta\right)}{\cos\left( \vartheta \right)}\nonumber\\ 
			=& \iu p r \tau + \frac{a\left(b_i - \iu p\rho\sigma\right)}{\sigma^2}\tau - \frac{2a}{\sigma^2} \ln \left(\cos\left(\frac{\iu \gamma\tau}{2}\right) - \sin\left(\frac{\iu\gamma\tau}{2}\right)\tan\left(\vartheta\right)\right)\nonumber\\ 
			=& \iu p r \tau + \frac{a\left(b_i - \iu p\rho\sigma\right)}{\sigma^2}\tau - \frac{2a}{\sigma^2} \ln \left(\cosh\left(\frac{\gamma\tau}{2}\right) - \iu\sinh\left(\frac{\gamma\tau}{2}\right)\frac{\iu\lambda}{\gamma}\right)\nonumber\\ 
			=& \iu p r \tau + \frac{a\left(b_i - \iu p\rho\sigma\right)}{\sigma^2}\tau - \frac{2a}{\sigma^2} \ln \left(\frac{e^{\frac{\gamma\tau}{2}} + e^{-\frac{\gamma\tau}{2}}}{2} + \frac{\lambda}{\gamma}\cdot\frac{e^{\frac{\gamma\tau}{2}} - e^{-\frac{\gamma\tau}{2}}}{2}\right)\nonumber\\ 
			=& \iu p r \tau + \frac{a\left(b_i - \iu p\rho\sigma\right)}{\sigma^2}\tau - \frac{2a}{\sigma^2} \ln e^{-\frac{\gamma\tau}{2}}\frac{\gamma\left(e^{\gamma\tau} + 1\right) + \lambda\left(e^{\gamma\tau} - 1\right)}{2\gamma}\nonumber\\ 
			=& \iu p r \tau + \frac{a\left(b_i - \iu\sigma \rho p+ \gamma\right)}{\sigma^2}\tau + \frac{2a}{\sigma^2} \ln \frac{2\gamma}{\left(\gamma + \lambda\right)e^{\gamma\tau}  + \gamma - \lambda}. 
		\end{align*} 
		Equation (\ref{ansatz2_beta_2}) may be simplified further as:
		\begin{align*} 
			B(t) =& \frac{\iu \gamma }{\sigma^2}\tan\left(\frac{\iu\gamma\tau}{2} + \vartheta\right) + \frac{b_i - \iu \sigma\rho p}{\sigma^2}
			= \frac{\iu \gamma }{\sigma^2}\left(\frac{\tan\left(\frac{\iu\gamma\tau}{2}\right) + \tan\left(\vartheta\right)}{1 - \tan\left(\frac{\iu\gamma\tau}{2}\right)\tan\left(\vartheta\right)}\right) + \frac{b_i - \iu \sigma\rho p}{\sigma^2}\nonumber\\ 
			=& \frac{\iu \gamma }{\sigma^2}\left(\frac{\iu\tanh\left(\frac{\gamma\tau}{2}\right) + \iu\frac{\lambda}{\gamma}}{1 - \iu\frac{\lambda}{\gamma}\iu\tanh\left(\frac{\gamma\tau}{2}\right)}\right) + \frac{b_i - \iu \sigma\rho p}{\sigma^2}
			= \frac{\iu\gamma }{\sigma^2}\left({\iu\frac{e^{\frac{\gamma\tau}{2}} - e^{-\frac{\gamma\tau}{2}}}{e^{\frac{\gamma\tau}{2}} + e^{-\frac{\gamma\tau}{2}}} + \iu\frac{\lambda}{\gamma}}\right)\left/\left({1 + \frac{\lambda}{\gamma}\frac{e^{\frac{\gamma\tau}{2}} - e^{-\frac{\gamma\tau}{2}}}{e^{\frac{\gamma\tau}{2}} + e^{-\frac{\gamma\tau}{2}}}}\right)\right. + \frac{b_i - \iu \sigma\rho p}{\sigma^2}\nonumber\\ 
			=& -\frac{\gamma }{\sigma^2}\left(\frac{e^{\frac{\gamma\tau}{2}} - e^{-\frac{\gamma\tau}{2}} + \frac{\lambda}{\gamma}\left(e^{\frac{\gamma\tau}{2}} + e^{-\frac{\gamma\tau}{2}}\right)}{e^{\frac{\gamma\tau}{2}} + e^{-\frac{\gamma\tau}{2}} + \frac{\lambda}{\gamma}\left(e^{\frac{\gamma\tau}{2}} - e^{-\frac{\gamma\tau}{2}}\right)}\right) + \frac{b_i - \iu \sigma\rho p}{\sigma^2}
			= -\frac{\gamma }{\sigma^2}\left(\frac{\gamma\left(e^{\gamma\tau} - 1\right) + \lambda\left(e^{\gamma\tau} + 1\right)}{\gamma\left(e^{\gamma\tau} + 1\right) + \lambda\left(e^{\gamma\tau} - 1\right)}\right) + \frac{b_i - \iu \sigma\rho p}{\sigma^2}\nonumber\\ 
			=& -\frac{\gamma }{\sigma^2}\left(\frac{\left(\gamma+\lambda\right)e^{\gamma\tau} + \lambda-\gamma}{\left(\gamma+\lambda\right)e^{\gamma\tau} + \gamma-\lambda}\right) + \frac{b_i - \iu \sigma\rho p}{\sigma^2}
			= -\frac{\gamma }{\sigma^2}\left(1 + \frac{2\left( \lambda-\gamma\right)}{\left(\gamma+\lambda\right)e^{\gamma\tau} + \gamma-\lambda}\right) + \frac{b_i - \iu \sigma\rho p}{\sigma^2}\nonumber\\ 
			=& \frac{2\gamma\left(\gamma-\lambda\right)}{\sigma^2\left(\left(\gamma+\lambda\right)e^{\gamma\tau} + \gamma-\lambda\right)} + \frac{b_i - \iu \sigma\rho p - \gamma}{\sigma^2}. 
		\end{align*} 
		Denoting $\tilde \zeta = 2\gamma/{\left( (\gamma + \lambda)\exp{(\gamma\tau)}  + \gamma - \lambda\right)}$, we have that the characteristic function is 
		\begin{align*} 
		\psi(p, q) =& \exp\left(A(t)x + B(t)v + C(t)\right)\\ 
			=& \exp\left(\iu p x + \left(\tfrac{2\gamma\left( \gamma-\lambda\right)}{\sigma^2\left(\left(\gamma+\lambda\right)e^{\gamma\tau} + \gamma-\lambda\right)} + \tfrac{b_i - \iu \sigma\rho p - \gamma}{\sigma^2}\right)v %
                        +\iu p r \tau + \tfrac{a\left(b_i - \iu\sigma \rho p+ \gamma\right)}{\sigma^2}\tau + \tfrac{2a}{\sigma^2} \ln \tfrac{2\gamma}{\left(\gamma + \lambda\right)e^{\gamma\tau}  + \gamma - \lambda}\right)\\ 
			=& \exp\left(\iu p \left(x +  r \tau\right) + \left(\frac{\left( \gamma-\lambda\right)\tilde \zeta}{\sigma^2} + \frac{\lambda - \gamma}{\sigma^2} + \iu q\right)v %
                        + \frac{\lambda + \gamma}{\sigma^2}a\tau + \iu qa\tau + \frac{2a}{\sigma^2} \ln \tilde \zeta\right)\\ 
			=& \exp\left(\iu p \left(x +  r \tau\right) + \iu q\left(v + a\tau\right) + \frac{ \gamma-\lambda}{\sigma^2}\left(\tilde \zeta - 1\right)v + \frac{\gamma+\lambda}{\sigma^2}a\tau +\frac{2a}{\sigma^2}\ln( \tilde{ \zeta}) \right). 
		\end{align*} 
		The logarithm term, $\ln ( \tilde{\zeta})$, may have a  discontinuity as $p$ increases. We can see that for $p\rightarrow\infty$, we have 
	        $\text{Re}(\gamma)\rightarrow\infty$,
                $\text{Im}(\gamma)\rightarrow\infty$, 
		and 
	        $\tilde \zeta \rightarrow 0$. 
		Though the value of $\tilde \zeta$ is bounded, the value of  of $\ln(\tilde{\zeta})$ will change very fast when $\tilde \zeta$ approaches $0$ and shifts phase eventually. To avoid the value of the logarithm term approaching either zero or infinity, we make the following  change of variables:
$$			\zeta = \frac{2\gamma}{\gamma + \lambda + (\gamma-\lambda)e^{-\gamma\tau}}; \quad 
			\tilde \zeta=\zeta e^{-\gamma\tau}; \quad
			\ln \tilde \zeta= -\gamma\tau + \ln \zeta ; \quad \text{and } \quad
			\frac{\gamma-\lambda}{\sigma^2}(\tilde \zeta -1)=\frac{\gamma+\lambda}{\sigma^2}(1 - \zeta).$$ 
		Therefore, our characteristic function is given as
		\begin{equation*} 
			\psi(p,q)=\exp\left(\iu p \left(x +  r \tau\right) + \iu q\left(v + a\tau\right) + \frac{ \gamma+\lambda}{\sigma^2}\left(1 - \zeta\right)v - \frac{\gamma-\lambda}{\sigma^2}a\tau +\frac{2a}{\sigma^2}\ln  \zeta\right). 
		\end{equation*} 
\noindent
This completes the derivation of the joint characteristic function as given in Theorem~\ref{thm_characteristic}.
	\end{proof} 

	\subsection{Proof of Proposition \ref{Prop_limiting_charac}}
        
	\begin{proof}\label{Pf_prop_limiting_charac} 
		We investigate the limiting behavior of the characteristic function given by (\ref{psi}) and the parameters $\gamma$, $\lambda$ and $\zeta$ as defined in equations (\ref{charac_params1})-(\ref{charac_params3}).  We have
		\begin{align*} 
			\lim\limits_{p\rightarrow\infty}\frac{\gamma}{p} =&  \sigma\sqrt{1-\rho^2} \sgn(p),\\ 
			\lim\limits_{p\rightarrow\infty}\frac{\lambda}{p} =& -\sigma\rho\iu, \mbox{ and}\\ 
			\lim\limits_{p\rightarrow\infty}\zeta =& 2\sqrt{1-\rho^2}\left(\sqrt{1-\rho^2} + \sgn(p)\rho\iu\right). 
		\end{align*} 
		Therefore,
		\begin{equation}
			\lim\limits_{p\rightarrow\infty}\ln\zeta = \ln\left(2\sqrt{1-\rho^2}\right) + \ln\left(\sqrt{1-\rho^2} + \sgn(p)\rho\iu\right). \label{ln_zeta} 
		\end{equation} 
		Further, note that
		\begin{align}\label{cc1} 
			\lim\limits_{p\rightarrow\infty}\frac{1}{p}\frac{ \gamma+\lambda}{\sigma^2}\left(1 - \zeta\right)v =& \lim\limits_{p\rightarrow\infty}\frac{\gamma+\lambda}{\sigma^2 p}\left(1-\frac{2\gamma}{\gamma + \lambda + (\gamma-\lambda)e^{-\gamma\tau}}\right) \nonumber \\
			&= \lim\limits_{p\rightarrow\infty}\frac{\gamma+\lambda}{\sigma^2 p}\left(\frac{\lambda -\gamma + (\gamma-\lambda)e^{-\gamma\tau}}{\gamma + \lambda + (\gamma-\lambda)e^{-\gamma\tau}}\right) 
			= \lim\limits_{p\rightarrow\infty} \frac{\lambda-\gamma}{\sigma^2 p} \nonumber \\
			&= -\frac{\sqrt{1-\rho^2}}{\sigma}v \sgn(p) - \frac{\rho }{\sigma}v\iu, 
		\end{align} 
		and 
		\begin{align}\label{cc2} 
			\lim\limits_{p\rightarrow\infty}\frac{1}{p}\frac{\gamma-\lambda}{\sigma^2}a\tau = \frac{\sqrt{1-\rho^2}}{\sigma}a\tau\sgn(p) - \frac{a\tau}{\sigma}\iu. 
		\end{align} 
		Let $\vartheta = \arcsin\left(\rho\sgn(p)\right)$ and transform (\ref{ln_zeta}) as 
		\begin{equation}\label{cc3} 
			\lim\limits_{p\rightarrow\infty}\ln\zeta(p) = \ln\left(2\sqrt{1-\rho^2}\right) + \vartheta \iu. 
		\end{equation} 
		Combining (\ref{cc1}), (\ref{cc2}) and (\ref{cc3}), we finalize the proof 
		\begin{equation*} 
			\lim\limits_{p\rightarrow\infty}\psi_i(p) \approx  A_\infty e^{\iu B_\infty} \exp\left(-\frac{\sqrt{1-\rho^2}}{\sigma}\left(v+a\tau\right)\left|p\right|\right). 
		\end{equation*} 
		where 
		\begin{align*} 
			A_\infty =& \left(4\left(1-\rho^2\right)\right)^{\frac{a}{\sigma^2}},\\ 
			B_\infty=& \frac{2a}{\sigma^2}\arcsin\left(\rho\sgn(p)\right) - \frac{\rho}{\sigma}\left(v + a\tau\sgn(p) \right) p. 
		\end{align*} 
This completes the proof of Proposition \ref{Prop_limiting_charac}.
	\end{proof}

	\subsection{Proof of Theorem \ref{thm_error_bound}} 
We next consider the truncation and discretization error bounds of the CFFT methods given in Theorem~\ref{thm_error_bound}.
	\begin{proof}\label{Pf_thm_error_bound} 
		We see that 
		\begin{equation*} 
			E_i(x) = \int_\mathbb{R} f(y) h_i(x-y)dy, 
		\end{equation*} 
		with $f(y)$ replaced by its Fourier expansion given in (\ref{F_seires}) 
		\begin{align}\label{Pi_expansion} 
			P_i(x) =& \int_\mathbb{R} \sum_{j=-\infty}^{\infty} F_j e^{-\iu j \frac{2\pi y}{L}} h_i(x-y) dy\nonumber\\ 
			=& \sum_{j=-\infty}^{\infty} F_j e^{-\iu j \frac{2\pi x}{L}} \int_\mathbb{R} e^{\iu j \frac{2\pi (x-y)}{L}} h_i(x-y) dy\nonumber\\ 
			=& \sum_{j=-\infty}^{\infty} F_j e^{-\iu j \frac{2\pi x}{L}} \int_\mathbb{R} e^{\iu j \frac{2\pi y}{L}} \phi_i(y) dy. 
		\end{align} 
		Replace the integral in equation (\ref{Pi_expansion}) by the kernel function (\ref{kernel}) 
		\begin{equation}\label{Pi_sum} 
			P_i(x) =\sum_{j=-\infty}^{\infty} F_j e^{-\iu j \frac{2\pi x}{L}} \psi_i\left(\frac{2\pi j}{L}\right). 
		\end{equation} 
		We truncate the infinite summation in equation (\ref{Pi_sum}) from $-\frac{N}{2}$ to $\frac{N}{2}-1$ 
		\begin{equation}\label{P_dot} 
			\dot P_i(x) = \sum_{j=-\frac{N}{2}}^{\frac{N}{2}-1} F_j e^{-\iu j \frac{2\pi x}{L}} \psi_i\left(\frac{2\pi j}{L}\right), 
		\end{equation} 
		and denote the truncation error as $e_{i,1}$ 
		\begin{align*} 
			\left|e_{i,1}\right| =& \left|P_i(x) - \dot P_i(x)\right|
			= \left|\sum_{j=-\frac{N}{2}-1}^{-\infty} F_j e^{-\iu j \frac{2\pi x}{L}} \psi_i\left(\frac{2\pi j}{L}\right) + \sum_{j=\frac{N}{2}}^{\infty} F_j e^{-\iu j \frac{2\pi x}{L}} \psi_i\left(\frac{2\pi j}{L}\right)\right|
			\leq \sum_{\left|j\right|=\frac{N}{2}}^{\infty} \left|F_j\right| \left|\psi_i\left(\frac{2\pi j}{L}\right)\right|.\nonumber 
		\end{align*} 
		By Proposition \ref{Prop_limiting_charac}, there exists a positive constant $\epsilon_{v,\tau}$ such that 
		\begin{equation}\label{bound1} 
			\left|\psi_i(p)\right| \leq \epsilon_{v,\tau} A_\infty \exp\left(-\frac{\sqrt{1-\rho^2}}{\sigma}\left(v+a\tau\right)\left|p\right|\right), \text{ for all } p. 
		\end{equation} 
		Denote 
		\begin{equation*} 
			D = \frac{\sqrt{1-\rho^2}}{\sigma}\left(v+a\tau\right). 
		\end{equation*} 
		Combining (\ref{bound1}) and (\ref{F_boundary}) yields 
		\begin{align*} 
			\left|e_{i,1}\right| \leq& \, 2 \epsilon_{v,\tau}\bar{f}\, A_\infty \sum_{j=\frac{N}{2}}^{\infty}  \exp\left({-D\left|\frac{2\pi j}{L}\right|}\right)  
			\leq  \, 2 \epsilon_{v,\tau}\bar{f}\, A_\infty \frac{L}{2\pi} \int_{\frac{\pi(N-2)}{L}}^{\infty}  \exp\left({-Du}\right) du\nonumber\\ 
			\leq& \frac{\epsilon_{v,\tau}\bar{f} L A_\infty}{\pi}  \exp{\left({-D \pi(N-2)}/{L}\right)}
			= \epsilon_1 \exp{(-\pi D N/L)},\nonumber 
		\end{align*} 
		where 
		\begin{equation*} 
			\epsilon_1 = \frac{L A_\infty e^{\frac{2\pi D}{L}}}{\pi D}\epsilon_{v,\tau}\bar{f}. 
		\end{equation*} 
		Next, we consider the discretization error arising from the DFT (\ref{tilde_P}) which is equivalent to the following calculation 
		\begin{equation*} 
			\tilde P_i(x) = \sum_{j=-\frac{N}{2}}^{\frac{N}{2}-1} \tilde F_j e^{-\iu j \frac{2\pi x}{L}} \psi_i\left(\frac{2\pi j}{L}\right), 
		\end{equation*} 
		by approximating the Fourier coefficients $F_j$ in (\ref{P_dot}) with 
		\begin{equation*} 
			\tilde F_j = \frac{\Delta x}{L}\sum_{k=0}^{N-1} f(x_k) e^{\iu j \frac{2\pi x_k}{L}}. 
		\end{equation*} 
		We denote the discretization error as $e_{i,2}$ 
		\begin{align}\label{dis_error} 
			\left|e_{i,2}\right| =& \left|\dot P_i(x) - \tilde P_i(x)\right|
			\leq  \sum_{j=-\frac{N}{2}}^{\frac{N}{2}-1} \left|F_j - \tilde F_j\right|\left|\psi_j\left(\frac{2\pi j}{L}\right)\right|. 
		\end{align} 
		Assuming that the discretization error of $\left|F_j - \tilde F_j\right|$ is of $O\left(N^{-m}\right)$, we can bound it with a positive bounding constant $\epsilon_L$ depending only on $L$ 
		\begin{equation}\label{F_dis} 
			\left|F_j - \tilde F_j\right| \leq \epsilon_L N^{-m}. 
		\end{equation} 
		It is easy to see that under the trapezoidal rule for $w_n$, we can apply $m\geq2$ in (\ref{F_dis}). Using the fact that $u_j=\frac{2\pi j}{L}$ for $j=-\frac{N}{2},\cdots,\frac{N}{2}-1$, we finalize the approximation in (\ref{dis_error}) as 
		\begin{align*} 
			\left|e_{i,2}\right| \leq& \sum_{j=-\frac{N}{2}}^{\frac{N}{2}-1} \left|F_j - \tilde F_j\right|\left|\psi_j\left(\frac{2\pi j}{L}\right)\right|
			\leq \, \epsilon_L  \epsilon_{v,\tau} A_\infty N^{-m} \sum_{j=-\frac{N}{2}}^{\frac{N}{2}-1} \exp\left(-D\left|\frac{2\pi j}{L}\right|\right)\nonumber\\ 
			\leq& \, \epsilon_L  \epsilon_{v,\tau} A_\infty N^{-m}  \frac{L}{2\pi}\int_{-\frac{N\pi}{L}}^{\frac{N\pi}{L}} \exp\left(-D\left|u\right|\right)du
			\leq \frac{\epsilon_L \epsilon_{v,\tau} L A_\infty N^{-m}}{\pi}\int_{0}^{\infty} \exp\left(-Du\right)du \\
			=& \frac{\epsilon_L \epsilon_{v,\tau} L A_\infty N^{-m}}{\pi D} 
			= \epsilon_2 N^{-m},
		\end{align*} 
		where 
		\begin{equation*} 
			\epsilon_2 = \frac{\epsilon_L \epsilon_{v,\tau} . L A_\infty}{\pi D}
		\end{equation*} 
		Therefore, the absolute error of the approximation of $P_i$ can be summarized as 
		\begin{align*} 
			\left|e_i\right| =& \left|P_i(x) - \tilde P_i(x)\right|
			\leq \left|P_i(x) - \dot P_i(x)\right| + \left|\dot P_i(x) - \tilde P_i(x)\right|\nonumber\\ 
			\leq& \left|e_{i,1}\right| + \left|e_{i,2}\right|
			\leq \epsilon_1 \exp{\left(-\pi D N / L \right)} + \epsilon_2 N^{-m}\nonumber, 
		\end{align*} 
		where the first component gives the upper bound of the truncation error and the second component gives the upper bound of the discretization error.  This completes the proof of Theorem \ref{thm_error_bound}.                
	\end{proof} 
	 
\end{document}